\documentclass[apj]{emulateapj}
\usepackage{multirow}

\makeatletter

\newenvironment{inlinefigure}{%
\def\@captype{figure}%
\noindent\begin{minipage}{0.999\linewidth}\begin{center}}
{\end{center}\end{minipage}\smallskip}
\makeatother

\shorttitle{Barger et al.}

\begin{document}

\title{Precise Identifications of Submillimeter Galaxies: Measuring the
History of Massive Star-Forming Galaxies to $z>5$\altaffilmark{1,2,3}}

\author{
A.~J.~Barger\altaffilmark{4,5,6}, 
W.-H.~Wang\altaffilmark{7},
L.~L.~Cowie\altaffilmark{6},
F.~N.~Owen\altaffilmark{8},
C.-C.~Chen\altaffilmark{6},
J.~P. Williams\altaffilmark{6}
}

\altaffiltext{1}{The Submillimeter Array is a joint project between the
Smithsonian Astrophysical Observatory and the Academia Sinica Institute
of Astronomy and Astrophysics and is funded by the Smithsonian Institution
and the Academia Sinica.}
\altaffiltext{2}{The National Radio Astronomy Observatory is a facility of the National Science Foundation operated under cooperative agreement by Associated Universities, Inc.}
\altaffiltext{3}{Herschel is an ESA space observatory with science instruments provided by European-led Principal Investigator consortia and with important participation from NASA.}
\altaffiltext{4}{Department of Astronomy, University of Wisconsin-Madison,
475 N. Charter Street, Madison, WI 53706.}
\altaffiltext{5}{Department of Physics and Astronomy, University of Hawaii,
2505 Correa Road, Honolulu, HI 96822.}
\altaffiltext{6}{Institute for Astronomy, University of Hawaii,
2680 Woodlawn Drive, Honolulu, HI 96822.}
\altaffiltext{7}{Academia Sinica Institute of Astronomy and Astrophysics, 
P.O. Box 23-141, Taipei 10617, Taiwan}
\altaffiltext{8}{National Radio Astronomy Observatory, P.O. Box O, 
Socorro, NM 87801.}

\slugcomment{The Astrophysical Journal, 761:89 (23pp), 2012 December 20}


\begin{abstract}

We carried out extremely sensitive Submillimeter Array (SMA) 340~GHz
(860~$\mu$m)
continuum imaging of a complete sample of SCUBA 850~$\mu$m sources
($>4\sigma$) with fluxes $>3$~mJy in the GOODS-N.
Using these data and new SCUBA-2 data,
we do not detect 4 of the 16 SCUBA sources, and we rule out the
original SCUBA fluxes at the $4\sigma$ level.
Three more resolve into multiple
fainter SMA galaxies, suggesting that our understanding of the most luminous
high-redshift dusty galaxies may not be as reliable as we thought.
10 of the 16 independent SMA sources have spectroscopic
redshifts (optical/infrared or CO) to $z=5.18$.
Using a new, ultradeep 20~cm image obtained with the
Karl G. Jansky Very Large Array (rms of 2.5~$\mu$Jy), we find that
all 16 of the SMA sources are detected at $>5\sigma$.
Using {\em Herschel\/} far-infrared (FIR) data, we show that the five 
isolated SMA sources with {\em Herschel\/} detections are well
described by an Arp~220 spectral energy distribution template in the
FIR. They also closely obey the local FIR-radio correlation,
a result that does not suffer from a radio bias.
We compute the contribution from the 16 SMA sources to the
universal star formation rate per comoving volume.
With individual star formation rates in the range
$700-5000$~M$_{\sun}$~yr$^{-1}$, they contribute
$\sim30$\% of the extinction-corrected ultraviolet
selected star formation rate density from $z=1$ to at least $z=5$.
Star formation histories determined from extinction-corrected
ultraviolet populations and from submillimeter galaxy
populations only partially overlap, due to the extreme ultraviolet
faintness of some submillimeter galaxies.

\end{abstract}


\section{Introduction}
\label{secintro}

The detection of the far-infrared (FIR) extragalactic background light 
(EBL) by {\em COBE\/} (e.g., Puget et al.\ 1996; Fixsen et al.\ 1998;
Hauser et al.\ 1998) was soon followed by the resolution 
of the bulk of the EBL at 850~$\mu$m
into discrete sources through deep 
submillimeter surveys with the Submillimeter Common-User Bolometer Array 
(SCUBA; Holland et al.\ 1999) on the 
single-dish James Clerk Maxwell Telescope (JCMT) 15~m
(e.g., Smail et al.\ 1997; Barger et al.\ 1998, 1999; Hughes et al.\ 1998;
Eales et al.\ 1999, 2000; Blain et al.\ 1999; Cowie et al.\ 2002; 
Scott et al.\ 2002; Webb et al.\ 2003; Borys et al.\ 2003;
Serjeant et al.\ 2003;
Wang et al.\ 2004; Pope et al.\ 2005; 
Coppin et al.\ 2006; Knudsen et al.\ 2008).
Millimeter continuum surveys were also undertaken
using the AzTEC (1.1~mm) bolometric camera (Wilson et al.\ 2008)
on both the JCMT (e.g., Scott et al.\ 2008; Perera et al.\ 2008; 
Austermann et al.\ 2010; Micha{\l}owski et al.\ 2012) 
and the Atacama Submillimeter Telescope Experiment (ASTE; Ezawa et al.\ 2004) 10~m
(e.g., Aretxaga et al.\ 2011; Scott et al.\ 2010, 2012; Yun et al.\ 2012)
and using the Max-Planck Millimeter Bolometer (MAMBO; Kreysa et al.\ 1998) 
array (1.2~mm) on the Institut de Radioastronomie Millim{\'e}trique (IRAM) 
30~m telescope (e.g., Dannerbauer et al.\ 2004; Gr{\`e}ve et al.\ 2004, 2008; 
Bertoldi et al.\ 2007; Lindner et al.\ 2011). 
Most recently, additional submillimeter surveys have been done at 
870~$\mu$m using the Large APEX Bolometer Camera 
(LABOCA; Siringo et al.\ 2009) on the Atacama Pathfinder 
Experiment (APEX; G{\"u}sten et al.\ 2006) 12~m telescope 
(e.g., Wei\ss\ et al.\ 2009; Johansson et al.\ 2011; Wardlow et al.\ 2011).

Through all of this work, we have learned that a large fraction of 
cosmic star formation is hidden by dust 
(e.g., Barger et al.\ 2000; Lagache et al.\ 2005; Chapman et al.\ 2005; 
Wang et al.\ 2006; Serjeant et al.\ 2008), 
and hence that the construction of a complete picture of galaxy 
evolution also requires an understanding of FIR and submillimeter 
galaxies (SMGs).
However, the challenge in studying the properties of this important
population of galaxies in detail has always been the low resolution 
of the FIR/submillimeter/millimeter data, which makes it difficult to 
identify the correct counterparts and to carry out follow-up 
observations.

One popular approach for pinpointing SMGs has been to use deep radio 
interferometric images. This approach takes advantage of the
well-known tight empirical correlation between non-thermal radio 
emission and thermal dust emission (e.g., Helou et al.\ 1985; Condon 1992).
Indeed, about 60\%$-$70\% of bright ($S_{850~\mu{\rm m}}\gtrsim5$~mJy) 
SMGs are found to have radio counterparts above 20~cm
flux limits of $\sim30~\mu$Jy (e.g., Barger et al.\ 2000; 
Ivison et al.\ 2002; Chapman et al.\ 2003b). With the
subarcsec positional accuracy of the radio data, it has
been possible to follow up a number of these bright SMGs 
spectroscopically.  Chapman et al.\ (2003a, 2005) found a
redshift distribution between $z\sim 1.5-3.5$ for these sources.

We note in passing that mid-infrared (MIR) data from {\em Spitzer\/} 
are also commonly used to identify counterparts to SMGs 
(e.g., Pope et al.\ 2006), but the resolution of these data is 
lower ($\sim2''$ resolution for IRAC; the 24~$\mu$m MIPS 
catalogs usually require IRAC source positions as priors) than in 
the radio. More recently, {\em Herschel\/}-PACS data (beam sizes of 
$6\farcs7$ and $11\farcs0$ at 100~$\mu$m and 160~$\mu$m, respectively)
have been used (e.g., Dannerbauer et al.\ 2010).  
Finally, hard X-ray observations with {\em Chandra\/}  
can also identify dusty active galactic nuclei (AGNs)  
(e.g., Bautz et al.\ 2000; Severgnini et al.\ 2000;
Barger et al.\ 2001a,b; Alexander et al.\ 2003a).

Unfortunately, there are several drawbacks to using radio emission to 
identify SMGs. First, with the current sensitivity of radio
interferometers ($5\sigma$ of $\sim20~\mu$Jy at 20~cm prior to the
upgrade of the Karl G. Jansky Very Large Array or VLA, 
e.g., Fomalont et al.\ 2006; 
Owen \& Morrison 2008; Morrison et al.\ 2010; Wold et al.\ 2012;
and now $5\sigma$ of $\sim12.5~\mu$Jy with the upgrade, e.g.,
F. Owen et al.\ 2013, in preparation; hereafter, Owen13), 
the radio-identified 
SMGs are mostly bright in the submillimeter, so their properties may 
not be representative of SMGs as a whole. Second, the radio 
flux drops at high redshifts due to the positive $K$-correction of
the radio synchrotron emission, while the submillimeter flux 
remains almost invariant over the redshift range $z\sim1-8$ 
(Blain \& Longair 1993) due to the negative $K$-correction of the
submillimeter thermal dust emission. Compton
cooling on the cosmic microwave background (CMB) radiation may also
significantly reduce the 20~cm emission in the highest
redshift sources (Condon 1992). Thus, the radio identification 
technique is biased against high-redshift sources. Finally, a number
of SMGs have been found to have more than one candidate radio 
counterpart (e.g., Ivison et al.\ 2002, 2007; Pope et al.\ 2006), making
identifying the correct counterpart to the SMG or determining 
whether there may be multiple counterparts difficult.

High-resolution interferometric imaging at the wavelength of the
original detection is clearly a more reliable way to identify the 
correct counterparts to SMGs. It is also a way to confirm that
an SMG is real rather than a false positive.
This is particularly important when counterparts at other wavelengths
cannot be found, since then the source could be interpreted as being
at a very high redshift when it is in fact not real.

Observations with the Submillimeter Array (SMA; Ho et al.\ 2004) 
have proven to be very useful for localizing the 
submillimeter emission from SCUBA sources
(e.g., Iono et al.\ 2006; Wang et al.\ 2007, 2011; 
Younger et al.\ 2008; Cowie et al.\ 2009; Hatsukade et al.\ 2010;
Knudsen et al.\ 2010; Chen et al.\ 2011),
as have millimeter observations with the IRAM Plateau de Bure 
Interferometer (PdBI)
using either continuum emission (e.g., Downes et al.\ 1999;
Dannerbauer et al.\ 2008) 
or CO molecular line emission 
(e.g., Daddi et al.\ 2009a,b; Walter et al.\ 2012).
Indeed, through the above studies of the SCUBA sources and 
other studies made using various telescope and instrument 
combinations (e.g., Capak et al.\ 2008, 2011; Schinnerer et al.\ 2008;
Coppin et al.\ 2009, 2010; Riechers et al.\ 2010;
Smol\v{c}i\'{c} et al.\ 2011),
the redshift distribution of SMGs has now been extended to $z>5$.

Moreover, the SMA observations have successfully discriminated
between multiple plausible counterparts identified at other wavebands.
For example, in several cases, only one of the multiple candidate 
radio counterparts was found to be the source of the submillimeter
emission (Younger et al.\ 2007, 2008, 2009; Hatsukade et al.\ 2010).

However, perhaps the most interesting and important result
to come out of the limited number of existing SMA interferometric 
localizations is the discovery that some of the brighter SCUBA sources
resolve into multiple, physically unrelated SMGs (Wang et al.\ 2011).
The Wang et al.\ results reinforce the notion that
cross-identifications of the dusty galaxies at any wavelength other 
than the submillimeter can be misleading, since such identifications 
are typically based on only one of the real counterparts.
Thus, the redshift distribution of the 
dusty star formation may need to be revised.

For the last few years, we have been observing with the SMA
the $>4\sigma$ sample of SCUBA-identified 
SMGs with 850~$\mu$m fluxes $>3$~mJy 
in the Great Observatories Origins Deep Survey-North (GOODS-N;
Giavalisco et al.\ 2004) presented by 
Wang et al.\ (2004; hereafter, W04; see their Table~1). 
There are 16 such sources listed in the W04 table.
Most of these sources are also in the 
SCUBA Super-map catalog of 
Pope et al.\ (2005; see their Table~A1). W04 used the naming
convention (prefix) GOODS~850, while P05 used the naming 
convention (prefix) GN for GOODS-N.
In addition, some of the sources in the Hubble Deep Field-North proper
(HDF-N; Williams et al.\ 1996) appeared in the 
Hughes et al.\ (1998) discovery
paper, where they used the naming convention (prefix) HDF~850.
For clarity and ease of comparison with the literature, where there
are overlaps, we always give all relevant names.

Note that the reason we imposed the $>3$~mJy flux limit is
because anything fainter than this is too hard to do with the 
sensitivity of the SMA. Applying this flux limit only 
removed GOODS 850-10 (GN~13 or HDF~850.4) from the $4\sigma$ sample
of W04, where it is listed as having a SCUBA flux of $2.6\pm0.5$~mJy.

Many recent analyses have focused on large samples of 
submillimeter or millimeter galaxies, which 
usually come at the expense of being heterogeneously selected,
not uniformly of high significance, and often without
high-resolution submillimeter or millimeter imaging. 
As a consequence, some of the sources may 
not be real or may have overestimated submillimeter fluxes,
while others may be multiples and/or have their counterparts 
misidentified. 
It is well-known in the literature that using lower significance 
thresholds is very dangerous due to the increased
contamination from flux boosting 
(e.g., Eddington 1913; Eales et al.\ 2000; Scott et al.\ 2002; Wang et al.\ 2004).
Moreover, as shown by Wang et al.\ (2011), multiple sources can
blend in single-dish observations to form a single, apparently more 
luminous source\footnotemark[1], 
the components of which may not even be physically related.

\footnotetext[1]{We note that after this paper was submitted, 
Karim et al.\ (2012) presented Atacama Large
Millimeter Array (ALMA) observations of a large sample of 870~$\mu$m selected
submillimeter sources from the LABOCA Extended {\em Chandra\/}
Deep South Submillimeter Survey (LESS). They found no
ALMA sources with fluxes $>9$~mJy.  They also found that all of the 
ALMA-observed $>12$~mJy 
sources  in the original LESS observations
were composed of emission from multiple, fainter SMGs. Both results are
more extreme than presented here:  we see three
single SMA sources with fluxes $>9$~mJy and confirm two of the three
$>12$~mJy sources in the original SCUBA observations as being single 
sources based on the SMA data.}

These recent analyses are orthogonal to the philosophy
of the present paper, which is to concentrate on an extremely 
well-understood sample of uniformly selected SMGs at the expense
of the sample being relatively small.
Here we analyze our targeted SMA sample as a whole.
We have already published a few of our SMA sources in previous 
work. In Wang et al.\ (2011), we 
presented GOODS~850-11 (GN~12) and GOODS~850-13 (GN~21), which 
we found to be composed of two and three distinct and unrelated 
sources, respectively. In Cowie et al.\ (2009), we presented 
GOODS~850-1 (GN14 or HDF~850.1), which continues to elude detection 
at optical/near-infrared (NIR) wavelengths but now has a redshift of 
$z=5.183$ from millimeter spectroscopy (Walter et al.\ 2012).
In Wang et al.\ (2007), we presented GOODS~850-5 (GN~10), which
is also undetected at optical/NIR wavelengths (Wang et al.\ 2009)
but now has a redshift of $z\sim4.05$ from millimeter spectroscopy 
(Daddi et al.\ 2009a).

Aiding our analysis, L.~Cowie et al.\ (2013, in preparation; 
hereafter, C13) 
recently obtained a 6~hr observation of the GOODS-N field in 
band 2 weather (225~GHz opacity $\sim0.05-0.08$) using SCUBA-2 
on the JCMT, providing a uniform sample over the field to 
$\sim8$~mJy ($4\sigma$).
SCUBA-2 is the most powerful camera for observing light 
at submillimeter wavelengths.
It covers 16 times the area of SCUBA and has a mapping speed that
is considerably faster than SCUBA at both 850~$\mu$m and
450~$\mu$m.  The arrays sample the sky in a way that is akin to CCDs 
or infrared cameras. The angular resolution of SCUBA-2 on the sky is 
a lot better than that of the space-based missions. For example,
at the {\em Herschel} satellite's longest wavelength (500~$\mu$m),
the beam FWHM size is $\sim35''$, whereas the beam FWHM sizes
of SCUBA-2 are $\sim7.5''$ and $\sim14''$ at 450~$\mu$m and 850~$\mu$m, 
respectively. The SCUBA-2 data provide an independent test of the 
reliability of the high-significance SCUBA samples in this field.

Also aiding our analysis, Owen13 recently obtained a 39~hr observation 
of the GOODS-N field at 20~cm with the upgraded VLA. 
This image reaches an rms of 2.5~$\mu$Jy. 
All of the galaxies in our SMA sample are detected 
in this image. Moreover, because of the high spatial resolution 
of both the submillimeter and radio images, the positions of the
galaxies, and the cross-identifications of the submillimeter
and radio sources, are unambiguous. We
use the combination of the submillimeter 
and radio measurements to estimate millimetric redshifts for the
galaxies without spectroscopic redshifts. We note that many
of the galaxies without spectroscopic redshifts are faint
in the optical/NIR, making the 
use of photometric redshifts challenging.
We also use the combined data to determine $q$ values ($q$ is a
measure of the logarithmic FIR/radio fux density ratio; 
e.g., Condon et al.\ 1991) in order to search for signs of 
evolution with FIR luminosity or redshift. Finally, we determine 
the radio luminosities for the galaxies, which we use to 
estimate the cosmic star formation history.

In Section~\ref{secobs}, we present our SMA observations and data
reduction. In Section~\ref{secsample}, we establish our SMA sample,
eliminating the sources in the original
SCUBA sample that we do not detect with the SMA and SCUBA-2.
In Section~\ref{secresults}, we analyze the sources in the sample, 
including determining the rate of multiplicity, the radio properties, 
the redshifts (spectroscopic and/or millimetric), and the optical/NIR 
properties. In Section~\ref{secclean}, we construct spectral energy 
distributions for the 5 SMGs in our SMA sample that are isolated 
and have {\em Herschel\/} detections. These SMGs have sufficient
wavelength coverage over the peak of the thermal dust spectrum to
constrain the quantity $q$. We include an additional source at 
$z=5.183$, whose 860~$\mu$m flux is close enough to the peak to
provide strong constraints on $q$, to look for
evolution in the high-redshift FIR-radio correlation to 
redshifts beyond 5.
In Section~\ref{secsfr}, we use our SMA sample to construct the 
star formation rate density over the redshift range $z=1-6$.
In Section~\ref{secsum}, we summarize our results.
We adopt the AB magnitude system for the optical and NIR photometry,
and we assume the Wilkinson Microwave
Anisotropy Probe cosmology of $H_0=70.5$~km~s$^{-1}$~Mpc$^{-1}$,
$\Omega_{\rm M}=0.27$, and $\Omega_\Lambda=0.73$ 
(Larson et al.\ 2011) throughout.

\section{SMA Observations and Data Reduction}
\label{secobs}

As we discussed in the introduction, we targeted with the SMA
W04's $>4\sigma$ sample of SCUBA-identified SMGs
with 850~$\mu$m fluxes $>3$~mJy in the GOODS-N.
In Table~\ref{tab1}, we give the details of our SMA observations
of each source. The only source in the sample of 16 that
we did not observe with the SMA was GOODS850-14, which is
not detected in the SCUBA-2 observations 
(see Section~\ref{secsample}).
The upgrade of the SMA to a 4~GHz bandwidth during the course
of our observing program considerably improved the continuum 
sensitivity and made calibrations with fainter quasars easier.
In Column~1, we give the W04 name for the SCUBA source, 
followed by the P05 and Hughes et al.\ (1998) names, where 
available; in Column~2, the date of the
observation; in Column~3, the system temperature (this contains
everything related to the sensitivity, including the opacity, 
the phase stability caused by air, and the instrument);
in Column~4, the exposure time on source; 
in Column~5, the number of antennas used;
in Column~6, the passband of the observations (we give the 
frequency center followed by the frequency range for each 
side band, where the latter is the same number for both side bands); 
in Column~7, the beam size;
in Column~8, the beam position angle;
in Column~9, the flux calibrator(s); 
in Column~10, the passband calibrator(s); and
in Column~11, the complex gain calibrators. The
observed frequencies range from 340 to 350 GHz
(850-880~$\mu$m), and we label them as 860~$\mu$m.

%
%
\begin{center}
\begin{deluxetable*}{lcccccccrcccc}
\tabletypesize{\scriptsize}
\tablecaption{SMA Observations}
\tablehead{\multicolumn{2}{c}{Name} & Date & T$_{\rm sys}$ & Exp. & \# & Passband\tablenotemark{a} & Beam & Beam & Flux & Passband & Complex Gain \\ 
GOODS & GN/HDF & & & & Ant. & & Size & PA &  
Calibrator(s) & Calibrator(s) & Calibrators \\
 & & & (K) & (hr) & & (GHz) & ($'' \times ''$)& (deg) & & & \\
\multicolumn{2}{c}{(1)} & (2) & (3) & (4) & (5) & (6) & (7) & (8) & (9) & (10) & (11) }
\startdata
850-1\tablenotemark{b} & 14/850.1 & 20080224 & 420 & 6.0 & 8 & 350, 2 & \multirow{2}{*}{$2.09\times 1.76$\tablenotemark{d}} & \multirow{2}{*}{60.2\tablenotemark{d}} & Ceres, Callisto & 3c111 & 1048+717, 1419+543 \cr
850-1\tablenotemark{c} & 14/850.1 & 20080421 & 610 & 5.9 & 7 & 340, 2 & & & Callisto & 3c454.3 & 1048+717, 1419+543 \cr
850-2 & 09 & 20090505 & 350 & 3.5 & 7 & 340, 2 & $2.08\times 1.28$ & 64.6 & Callisto & 3c273 & 1058+812, 1642+689 \cr
850-3 & 06 & 20110424 & 340 & 4.6 & 8 & 342, 4 & $2.22\times 2.12$ & -66.2 & Neptune & 3c454.3 & 1153+495, 1642+689 \cr
850-4 & \nodata & 20070215 & 630 & 6.9 & 7 & 340, 2 & $2.32\times 1.33$ & 49.6 & Callisto & 3c84 & 1048+717, 1419+543 \cr
850-5 & 10 & 20070123 & 380 & 6.6 & 8 & 340, 2 & $2.42\times 2.19$ & 24.3 & Callisto & 3c84 & 1048+717, 1419+543 \cr
850-6 & \nodata & 20110325 & 430 & 8.1 & 8 & 342, 4 & $2.25\times 2.00$ & 52.7 & Neptune & 3c84 & 1153+495, 1642+689 \cr
850-7 & 04 & 20100424 & 350 & 6.1 & 7 & 342, 4 & $2.42\times 2.07$ & -6.1 & Neptune & 3c454.3 & 0958+655, 1642+689 \cr
850-8 & \nodata & 20120505 & 380 & 6.0 & 6 & 342, 4 & $1.98\times 1.84$ & -64.4 & Neptune & 2202+422 & 1153+495, 1642+689 \cr
850-9 & 19 & 20100419 & 410 & 5.8 & 7 & 342, 4 & $2.59\times 2.07$ & 27.4 & Titan & 3c454.3 & 0958+655, 1642+689 \cr
850-11 & 12 & 20091230 & 380 & 4.8 & 7 & 340, 4 & $2.80\times 1.96$ & 35.0 & Titan & 3c273 & 0958+655, 1642+689 \cr
850-12 & 15/850.2 & 20100417 & 410 & 4.8 & 7 & 342, 4 & $2.38\times 2.09$ & 13.9 & Mars & 3c454.3, 3c84 & 0958+655, 1642+689 \cr
850-13 & 21 & 20091231 & 380 & 5.4 & 8 & 340, 4 & $2.27\times 2.03$ & 13.6 & Titan & 3c273 & 0958+655, 1642+689 \cr
850-15 & 07 & 20100425 & 380 & 6.0 & 7 & 342, 4 & $2.44\times 1.99$ & 4.7 & Neptune & 3c454.3 & 0958+655, 1642+689 \cr
850-16 & 16 & 20110424 & 340 & 2.6 & 8 & 342, 4 & $2.25\times 2.00$ & 52.7 & Neptune & 3c454.3 & 1153+495, 1642+689 \cr
850-17 & \nodata & 20120501 & 270 & 5.6 & 6 & 342, 4 & $2.48\times 2.02$ & 63.9 & Titan, Mars & 2202+422 & 1153+495, 1642+689
\enddata
\label{tab1}
\tablenotetext{a}{The first number is the frequency center, and the second
number is the frequency range for each side band;
it is the same number for both side bands.}
\tablenotetext{b}{First track.}
\tablenotetext{c}{Second track.}
\tablenotetext{d}{Each night individually has a different beam size and PA, but
since we never use the data separately, we quote here the combined beam size 
and PA for the two nights.}
\end{deluxetable*}
\end{center}

We performed the calibration and data inspection using the
IDL-based Caltech package MIR modified for the SMA. We
generated continuum data by averaging the spectral channels
after doing the passband phase calibration. We used both gain 
calibrators to derive gain curves. For consistency checks,
we compared these results with those obtained by adopting 
just one calibrator. We did not find any systematic differences. 
We computed the fluxes using calibrators observed on the same day
and under similar conditions (time, hour angle, and elevation).
The flux calibration error is typically 
within $\sim10$\% with this method.
We exported the calibrated interferometric visibility data
to the package MIRIAD for subsequent imaging and analysis.
We weighted the visibility data inversely proportional to the 
system temperature and Fourier transformed them to form images.
We also applied the ``robust weighting'' of Briggs (1995), with
a robust parameter of 1.0, to obtain a better balance between
beam size and signal-to-noise ratio (S/N). 
We CLEANed the images
around detected sources to approximately 1.5 times the
noise level to remove the effects of sidelobes. 
(The results are not sensitive to choosing a slightly deeper CLEANing
level, such as 1.0 times the noise.) 
We typically achieved rms $\sim1.2-1.5$ mJy in a night with the old
2~GHz bandwidth and rms $\sim0.7-0.9$ mJy in a night with the new 4~GHz 
bandwidth. We corrected the images for the SMA primary beam response. 
All of the SMA fluxes and flux errors that we quote are primary-beam 
corrected. We measured source positions and fluxes by fitting the 
images with point-source models using the MIRIAD IMFIT routine.

\section{SMA Sample}
\label{secsample}

C13 used their new SCUBA-2 observations of
the GOODS-N field (8~mJy sensitivity at $4\sigma$)
to test the reliability of the $>4\sigma$ SCUBA 
samples of W04 and P05.
The SCUBA-2 data rule out at the $>4\sigma$ level
three sources found only in the
P05 sample (GN02, GN03, GN08), two sources found only
in the W04 sample (GOODS~850-4 and GOODS~850-14), and
one source found in both samples (GOODS~850-16 or GN16).
We had already observed both GOODS~850-4 and GOODS~850-16
(GN16) with the SMA and had not detected either source.
Thus, the SMA and SCUBA-2 observations independently 
show that these sources are either spurious or are much
fainter than the original SCUBA fluxes suggested.
Although the SCUBA-2 data did not rule out at the $>4\sigma$
level GOODS~850-8, we did not detect this source with the SMA, 
so it also appears to be problematic.
We therefore remove the sources GOODS~850-4, GOODS~850-8,
GOODS~850-14, and GOODS~850-16 (GN16) from further consideration.

In Table 2, we give the names (Column 1) and J2000 right ascensions 
(Column 2) and declinations (Column 3), as measured from the SCUBA 
observations, for all of the unconfirmed sources.  We also provide the W04 
SCUBA (Column 4), P05 SCUBA (Column 5), C13 SCUBA-2 (Column 6), 
and SMA (where available; Column 7) fluxes and $1\sigma$ errors.

%
%
\begin{center}
\begin{deluxetable*}{lcllcccc}
\renewcommand\baselinestretch{1.0}
\tablecaption{Unconfirmed SCUBA Sources}
\tabletypesize{\scriptsize}
\tablehead{\multicolumn{2}{c}{Name} & SCUBA R.A.\tablenotemark{a} & SCUBA Decl.\tablenotemark{a} & W04 SCUBA & P05 SCUBA & C13 SCUBA-2  & SMA \\ 
GOODS & GN/HDF & J2000.0 & J2000.0 & 850~$\mu$m & 850~$\mu$m & 850~$\mu$m & 860~$\mu$m \\
& & ($^{\rm h}~^{\rm m}~^{\rm s}$) & ($^\circ~'~''$) & (mJy) & (mJy) & (mJy) & (mJy) \\
\multicolumn{2}{c}{(1)} & (2) & (3) & (4)  & (5) & (6) & (7)}
\startdata
850-4 & \nodata & 12 36 37.05 & 62 12 08.45 & $8.62\pm1.27$ & \nodata & -0.4$\pm 2.0$ & $<5\, (3\sigma)$ \cr
850-8 & \nodata & 12 36 06.30 & 62 12 47.05 & $8.13\pm 1.40$ & \nodata & $3.8\pm 2.1$ & $<3.9\, (3\sigma)$ \cr
850-14 & \nodata & 12 36 23.45 & 62 13 16.33 & $10.46\pm2.32$ & \nodata & $2.8\pm 2.0$ & \nodata \cr
850-16 & 16 & 12 37 00.05 & 62 09 15.48 & $12.45\pm2.88$ & $9.0\pm2.1$ & $0.5\pm 2.1$ & $<2\, (3\sigma)$ \cr
\nodata & 02 & 12 36 07.7 & 62 11 47 & \nodata & $16.2\pm 4.1$ & -1.0$\pm 2.0$ & \nodata \cr
\nodata & 03 & 12 36 08.9 & 62 12 53 & \nodata & $16.8\pm 4.0$ & $0.7\pm 2.1$ & \nodata \cr
\nodata & 08 & 12 36 22.2 & 62 12 56 & \nodata & $12.5\pm 2.7$ & -1.4$\pm 2.1$ & \nodata
\enddata
\label{tab2}
\tablenotetext{a}{Where the source has a GOODS 850 identification, we give the right ascension and 
declination measurements from W04. Where the source has only a GN identification, we give the right
ascension and declination measurements from P05.
}
\end{deluxetable*}
\end{center}

In Table~\ref{tab3}, we give the names (Column 1) of the 12 
remaining sources. In cases where we now have 
multiple SMA detections, we have added an ``a", ``b", or ``c" 
to the name to distinguish between them.  
Thus, the sample is actually composed of 16 independent sources rather than 12.
{\em Hereafter, we refer to these 16 sources as our SMA sample.\/}
In Columns~2 and 3, we give the J2000 right ascensions and 
declinations for each source as measured from the SMA observations.
In Column~4, we list the W04 SCUBA fluxes and $1\sigma$ errors;
in Column~5, the P05 SCUBA fluxes and $1\sigma$ errors;
in Column~6, the C13 SCUBA-2 fluxes and $1\sigma$ errors; and
in Column~7, the SMA fluxes and $1\sigma$ errors.

%
%
\begin{center}
\begin{deluxetable*}{lccccccc}
\renewcommand\baselinestretch{1.0}
\tablewidth{0pt}
\tablecaption{Submillimeter Properties of the SMA Sample}
\tabletypesize{\scriptsize}
\tablehead{\multicolumn{2}{c}{Name} & SMA R.A. & SMA Decl. & W04 SCUBA & P05 SCUBA & C13 SCUBA-2 & SMA \\ 
GOODS & GN/HDF & J2000.0 & J2000.0 & 850~$\mu$m & 850~$\mu$m & 850~$\mu$m & 860~$\mu$m \\
& & (deg) & (deg) & (mJy) & (mJy) & (mJy) & (mJy) \\
\multicolumn{2}{c}{(1)} & (2) & (3) & (4)  & (5) & (6) & (7)}
\startdata
850-1 & 14/850.1 & 189.216614 & 62.207165 & $5.1\pm 0.5$ & $5.9\pm 0.3$ & $8.0\pm 2.0$ & $7.8\pm 1.0$ \cr
850-2 & 09 & 189.092117 & 62.271030 & $10.3\pm 1.2$ & $8.9\pm 1.0$ & $6.1\pm 2.1$ & $9.3\pm 1.4$ \cr
850-3 & 06 & 189.076370 & 62.264111 & $7.7\pm 1.0$ & $7.5\pm 0.9$ & $7.2\pm 2.1$ & $7.2\pm 0.7$ \cr
850-5 & 10 & 189.139374 & 62.235752 & $12.9\pm 2.1$ & $11.3\pm 1.6$ & $10.5\pm 2.0$ & $12.0\pm 1.4$ \cr
850-6 & \nodata & 189.378326 & 62.216389 & $13.6\pm 2.3$ & \nodata & $14.9\pm 2.1$ & $14.9\pm 0.9$ \cr
850-7 & 04 & 189.067123 & 62.253834 & $6.2\pm 1.0$ &  $5.1\pm 1.0$ &  $7.9\pm 2.1$ & $3.4\pm 0.6$ \cr
850-9 & 19 & 189.280045 & 62.235638 & $7.1\pm 1.2$ & $10.7\pm 2.7$ & $9.2\pm 2.0$ & $7.1\pm 1.4$ \cr
850-11 & 12 & & & $10.8\pm 2.2$ & $8.6\pm 1.4$ & $4.3\pm 2.0$ & \cr
\quad\quad\quad a & & 189.192047 & 62.246830 & & & & $4.2\pm 0.8$ \cr
\quad\quad\quad b & & 189.183243 & 62.247417 & & & & $5.3\pm 1.1$ \cr
850-12 & 15/850.2 & 189.233002 & 62.200531 & $3.3\pm 0.7$ & $3.7\pm 0.4$ & $4.2\pm2.0$ &  $4.5\pm0.8$ \cr
850-13 & 21 & & & $7.0\pm 1.5$ & $5.7\pm 1.4$ & $2.2\pm 2.0$ & \cr
\quad\quad\quad a & & 189.308472 & 62.199001 & & & &  $3.2\pm 0.9$ \cr
\quad\quad\quad b & & 189.309433 & 62.202248 & & & & $4.1\pm  0.7$ \cr
\quad\quad\quad c & & 189.300003 & 62.203415 & & & & $5.3\pm 0.9$ \cr
850-15 & 07 & & & $8.7\pm 2.0$ & $8.9\pm 1.5$ & $0.3\pm 2.1$ & \cr
\quad\quad\quad a & & 189.087921 & 62.285999 & & & & $3.4\pm 0.6$ \cr
\quad\quad\quad b & & 189.088745 & 62.285883 & & & & $3.5\pm 0.7$ \cr
850-17 & \nodata & 189.120160 & 62.179242 & $5.7\pm 1.4$ & \nodata & $-0.1\pm 2.1$ & $7.7\pm 0.9$
\enddata
\label{tab3}
\end{deluxetable*}
\end{center}

We tested the astrometric accuracy of the SMA observations 
relative to the GOODS-N 20~cm sample of Owen13.
We find that they are in perfect astrometric agreement.
The dispersion in the position offsets is $0\farcs5$.
The two largest offsets are GOODS~850-12 at $1\farcs3$ and 
GOODS~850-15a at $0\farcs86$. 

We tabulate the multiwavelength properties of the SMA sample in
Table~\ref{tab4}, and we present the SMA and multiwavelength
images in Figure~A1 of the Appendix.
In Column~1 of Table~\ref{tab4}, we give the names of the SMA sources.
In Column~2, we give the $3''$ diameter aperture $K_s$ magnitudes 
corrected to total magnitudes. We measured these at the SMA 
positions for the isolated galaxies using the $K_s$ image of 
Wang et al.\ (2010). We do not provide measurements for GOODS~850-1 
(GN14 or HDF850.1), GOODS~850-12 (GN15 or HDF850.2), or GOODS~850-17, 
where there are nearby bright galaxies.

In Column~3, we give the $2''$ diameter aperture {\em HST\/} WFC3
F140W magnitudes corrected to total magnitudes. We measured
these at the SMA positions using the 
F140W images from the archival {\em HST\/} slitless WFC3/G141 IR 
grism survey obtained by B.~Weiner (Cycle~17, Proposal 11600). 
The only exceptions are GOODS~850-1, 
where we used a $1''$ diameter aperture to minimize contamination 
from the neighboring galaxies, GOODS~850-15a and 15b, where 
there is no F140W imaging, and GOODS~850-17, where the flux of
the source is too contaminated by the neighboring galaxy.

We searched a $1''$ radius around each SMA position to find
optical counterparts in the ACS catalog
of Giavalisco et al.\ (2004). 
We give the magnitudes of these matches in Columns~$4-7$.
Note that GOODS 850-2 has a negative flux in the $F140W$ band. 

We measured the 20~cm fluxes of the radio sources corresponding
to the individual SMA sources in the 39~hr VLA image of Owen13.
In Table~\ref{tab5}, we give the details of these measurements.
In Column~1, we give the names of the SMA sources; 
in Columns~2 and 3, the SMA J2000 right ascensions and declinations; 
in Columns~4 and 5, the VLA right ascensions and declinations; 
in Column~6, for the sources that are not resolved,
the upper limits on the sizes of the sources; 
in Column~7, the peak radio fluxes; 
in Columns~$8-11$, for the sources that are resolved, 
the total radio fluxes, major axes, minor axes, and position 
angles for the Gaussian fits.

\begin{deluxetable*}{lcccccccccc}
\tablewidth{0pt}
\tablecaption{Multiwavelength Properties of the SMA Sample}
\tabletypesize{\scriptsize}
\tablehead{
Name & $K_s$ & F140W & F850LP & F775W & F606W & F450W & 20~cm & $0.5-2$~keV & $2-8$~keV & $z_{spec}$ $z_{milli}$\tablenotemark{a} \\ 
GOODS (GN/HDF) & (AB) & (AB) & (AB) & (AB) & (AB) & (AB) & ($\mu$Jy) & \multicolumn{2}{c}{($10^{-16}$~erg~cm$^{-2}$~s$^{-1}$)} & \\
(1) & (2) & (3) & (4) & (5) & (6) & (7) & (8) & (9) & (10) & (11) (12)
}
\startdata
850-1 (14/850.1) & \nodata & 25.77  & \nodata & \nodata & \nodata & \nodata & 12.3 (17)\tablenotemark{b} & \nodata & \nodata & 5.183\tablenotemark{c} (4.6)\cr
850-2 (09) & 23.92 & -28.19 & \nodata & \nodata & \nodata & \nodata  & 17.8 & \nodata & \nodata & \nodata 4.3\cr
850-3 (06) & 21.96 & 23.15 & 26.19 & 27.35 & 28.10 & 29.28  & 161.9 & \nodata & \nodata & 2.000\tablenotemark{d} (1.8)\cr
850-5 (10) & 25.81 & 26.54 & \nodata & \nodata & \nodata & \nodata  & 32.2 & \nodata & \nodata & 4.0424\tablenotemark{e} (3.9)\cr
850-6 & 22.92 & 24.14 & 25.63 & 25.98 & 25.78 & 25.99 & 117.4 & \nodata & \nodata & \nodata 2.7\cr
850-7\tablenotemark{f} (04) & 22.24 & 23.55 & 25.60 & 25.55 & 38.6 & 27.51 & 38.6 & 1.80 & 7.97 & 2.578\tablenotemark{g} (2.3)\cr
850-9\tablenotemark{f} (19) & 21.43 & 23.01 & 24.99 & 25.39 & 25.89 & 26.90  & 25.9 & 0.921 & 9.11 & 2.490\tablenotemark{h} (3.5)\cr
850-11 (12) &  & &  &  &  &  &  &  &  & \cr
\quad\quad\quad a\tablenotemark{f} & 23.25 & 24.23 & 26.02 & 26.21 & 26.73 & 28.44 & 101.8 & -0.339 & -1.38 & \nodata 1.7\cr
\quad\quad\quad b & 21.55 & 22.49 & 23.93 & 24.21 & 24.83 & 25.31 & 30.5 & \nodata & \nodata & 2.095\tablenotemark{i} (3.0)\cr
850-12 (15/850.2) & \nodata & 23.89 & 24.27 & 24.59 & 24.78 & 25.16  & 19.0  &  1.26 &  2.70 & 2.737\tablenotemark{j} (3.3)\cr
850-13 (21) & &  &  &  &  & &  & & & \cr
\quad\quad\quad a & 24.70 & 26.04 & \nodata & \nodata & \nodata & \nodata & 21.7 (18)\tablenotemark{k} & \nodata & \nodata & \nodata 2.8\cr
\quad\quad\quad b & 21.10 & 22.32 & 23.57 & 23.92 & 24.31 & 24.76 & 23.4 (15)\tablenotemark{k} & 2.21 & 5.24 & 3.157\tablenotemark{l} (3.0)\cr
\quad\quad\quad c & 22.10 & 23.88 & 25.69 & 25.94 & 26.73 & 27.45 & 32.5 & 0.293 & 3.74 & 2.914\tablenotemark{m} (2.9)\cr
850-15\tablenotemark{f} (07) & & & & &  &  &  & &  \cr
\quad\quad\quad a & 21.77 & \nodata & 23.57 & 23.92 & 24.34 & 24.67 & 43.2 & \nodata & \nodata & 1.992\tablenotemark{n} (2.2) \cr
\quad\quad\quad b & 21.68 & \nodata & 26.81 & 27.83 & 28.68 & 28.41 & 146.9 & \nodata & \nodata & \nodata 1.4\cr
850-17\tablenotemark{f} & \nodata & \nodata & \nodata & \nodata & \nodata & \nodata & 48.5\tablenotemark{o} & \nodata & \nodata & \nodata 2.8
\enddata
\label{tab4}

\tablenotetext{a}{Where a spectroscopic redshift already exists
for the source, we put the millimetric redshift in parentheses.}

\tablenotetext{b}{The quantity in brackets is the previous measurement
from Cowie et al.\ (2009).} 

\tablenotetext{c}{The IRAM PdBI CO[5-4], CO[6-5], and [CII] redshift of 
$z=5.183$ is from Walter et al.\ (2012).}

\tablenotetext{d}{The Pope et al.\ (2008) {\em Spitzer\/} IRS redshift
of $z=2.00\pm0.03$ is inconsistent with the optical redshift of 
$z=1.865$ from Chapman et al.\ (2005). Bothwell et al.\ (2010) found an
IRAM PdBI CO[4-3] redshift of $z=1.999\pm0.001$ for the source, which 
they called HDF~132.}

\tablenotetext{e}{The IRAM PdBI CO[4-3] redshift of $z=4.0424$ is from 
Daddi et al.\ (2009a).}

\tablenotetext{f}{These sources have close pairs of radio sources.}

\tablenotetext{g}{The brighter radio source has an optical 
redshift of $z=2.578$ from Chapman et al.\ (2005). 
The fainter radio source lacks an optical counterpart. 
The counterpart to the brighter radio source, which
is at the position of the SMA source, is a chain-like 
object with three components. The magnitudes in the table 
correspond to the reddest component in this structure, which
is closest to the SMA and VLA positions. If the components 
are not physically related, then the redshift 
may not correspond to the submillimeter galaxy.}

\tablenotetext{h}{The two equally bright radio sources have 
optical redshifts of $z=2.484$ from Chapman et al.\ (2005). 
The NIR redshift of $z=2.490$ is from Swinbank et al.\ (2004).
The {\em Spitzer\/} IRS redshift of $z=2.48\pm0.03$ is from
Pope et al.\ (2008).}

\tablenotetext{i}{The optical redshift of $z=2.095$ is from 
Reddy et al.\ (2006).}

\tablenotetext{j}{The optical redshift of $z=2.737$ is from
Barger et al.\ (2008).}

\tablenotetext{k}{The quantities in brackets are the previous 
measurements from Wang et al.\ (2011).}

\tablenotetext{l}{The optical redshift of $z=3.157$ is from 
Barger et al.\ (2008).}

\tablenotetext{m}{The optical redshift of $z=2.914$ is from
Chapman et al.\ (2005).}

\tablenotetext{n}{The optical redshift of $z=1.99$ for the 
brighter optical but fainter radio source of the GOODS 850-15a/850-15b pair
is from Chapman et al.\ (2004).
The NIR redshift of $z=1.992$ for the same source is from 
Swinbank et al.\ (2004).} 

\tablenotetext{o}{The neighboring radio/X-ray source previously
identified as the correct counterpart to this submillimeter galaxy
has an optical redshift of $z=1.013$ from Barger et al.\ (2008). 
We now know the other radio source is the correct counterpart 
to this submillimeter galaxy.}

\end{deluxetable*}

\begin{deluxetable*}{cccccccccccc}
\renewcommand\baselinestretch{1.0}
\tablecaption{Radio Measurements}
\tabletypesize{\scriptsize}
\tablehead{
\multicolumn{2}{c}{Name} & SMA R.A. & SMA Decl. & VLA R.A. & VLA Decl. & Upper & Peak Radio & Total Radio & Major & Minor & P.A. \\
GOODS & GN/HDF & J2000.0 & J2000.0 & J2000.0 & J2000.0 & Limit & Flux & Flux & Axis & Axis & \\
& & ($^{\rm h}~^{\rm m}~^{\rm s}$) & ($^\circ~'~''$) & ($^{\rm h}~^{\rm m}~^{\rm s}$) & ($^\circ~'~''$) & ($''$) & ($\mu$Jy/beam) & ($\mu$Jy) & ($''$) & ($''$) & (deg) \\
\multicolumn{2}{c}{(1)} & (2) & (3) & (4) & (5) & (6) & (7) & (8) & (9) & (10) & (11)
}
\startdata
850-1 & 14/850.1 & 12 36 51.98 & 62 12 25.8 & 12 36 52.04 & 62 12 25.9 & $<1.5$ & $12.3\pm2.4$ & \nodata & \nodata & \nodata & \nodata \cr
850-2 & 09 & 12 36 22.11 & 62 16 15.7 & 12 36 22.11 & 62 16 15.9 & $<1.6$ & $17.8\pm2.5$ & \nodata & \nodata & \nodata & \nodata \cr
850-3 & 06 & 12 36 18.33 & 62 15 50.8 & 12 36 18.35 & 62 15 50.6 & \nodata & $155.0\pm2.7$ & $161.9\pm 4.8$ & 0.5 & 0.4 & $129\pm21$ \cr
850-5 & 10 & 12 36 33.45 & 62 14 08.7 & 12 36 33.42 & 62 14 08.5 & \nodata & $20.6\pm2.3$ & $32.2\pm5.5$ & 1.9 & 0 & $72\pm8$ \cr
850-6 & \nodata & 12 37 30.80 & 62 12 59.0 & 12 37 30.81 & 62 12 58.8 & $<0.6$ & $117.4\pm2.7$ & \nodata & \nodata & \nodata & \nodata \cr
850-7 & 04 & 12 36 16.11 & 62 15 13.8 & 12 36 16.11 & 62 15 13.7 & $<1.1$ & $38.6\pm2.6$ & \nodata & \nodata & \nodata & \nodata \cr
\quad\quad comp & \nodata & \nodata & \nodata & 12 36 15.84 & 62 15 15.6 & $<0.4$ & $28.7\pm2.6$ & \nodata & \nodata & \nodata & \nodata \cr
850-9 & 19 & 12 37 07.21 & 62 14 08.3 & 12 36 16.11 & 62 15 13.7 & $<1.0$ & $25.9\pm2.5$ & \nodata & \nodata & \nodata & \nodata \cr
\quad\quad comp & \nodata & \nodata & \nodata & 12 37 07.60 & 62 14 09.6 & $<1.0$ & $28.3\pm2.5$ & \nodata & \nodata & \nodata & \nodata \cr
850-11 & 12 & & & & & & & & & \cr
\quad\quad\quad a & & 12 36 46.09 & 62 14 48.6 & 12 36 46.08 & 62 14 48.6 & $<0.7$ & $101.8\pm2.4$ & \nodata & \nodata & \nodata & \nodata \cr
\quad\quad comp & \nodata & \nodata & \nodata & 12 36 46.81 & 62 14 45.5 & \nodata & $42.4\pm2.3$ & $68.9\pm5.7$ & 2.1 & 0 & $133\pm5$ \cr
\quad\quad\quad b & & 12 36 43.98 & 62 14 50.7 & 12 36 44.03 & 62 14 50.6 & $<0.6$ & $30.5\pm2.4$ & \nodata & \nodata & \nodata & \nodata \cr
850-12 & 15/850.2 & 12 36 55.92 & 62 12 01.9 & 12 36 55.80 & 62 12 00.9 & \nodata & $13.1\pm2.4$ & $19.0\pm5.4$ & 1.6 & 0.3 & $34\pm12$ \cr
850-13 & 21 & & & & & & & & & \cr
\quad\quad\quad a & & 12 37 14.03 & 62 11 56.40 & 12 37 14.05 & 62 11 56.6 & $<1.1$ & $21.7\pm2.5$ & \nodata & \nodata & \nodata & \nodata \cr
\quad\quad\quad b & & 12 37 14.26 & 62 12 08.1 & 12 37 14.29 & 62 12 08.5 & $<1.3$ & $23.4\pm2.5$ & \nodata & \nodata & \nodata & \nodata \cr
\quad\quad\quad c & & 12 37 12.00 & 62 12 12.3 & 12 37 12.06 & 62 12 11.9 & $<0.7$ & $32.5\pm2.4$ & \nodata & \nodata & \nodata & \nodata \cr
850-15 & 07 & & & & & & & & & \cr
\quad\quad\quad a & & 12 36 21.10 & 62 17 09.6 & 12 36 20.98 & 62 17 09.8 & $<1.0$ & $43.2\pm2.6$ & \nodata & \nodata & \nodata & \nodata \cr
\quad\quad\quad b & & 12 36 21.30 & 62 17 09.2 & 12 36 21.28 & 62 17 08.4 & $<0.6$ & $146.9\pm2.6$ & \nodata & \nodata & \nodata & \nodata \cr
850-17\tablenotemark{a} & \nodata & 12 36 28.84 & 62 10 45.3 & 12 36 28.90 & 62 10 45.3 & \nodata & $35.0\pm3.9$ & $48.5\pm9.4$ & 0.7 & 0.4 & $73\pm40$ \cr
\quad\quad comp & \nodata & \nodata & \nodata & 12 36 29.16 & 62 10 46.0 & \nodata & $36.5\pm3.9$ & $48.1\pm8.2$ & 1.0 & 0 & $30\pm12$ \cr
\enddata
\label{tab5}
\tablenotetext{a}{For this source, we use the Gaussian fit from the
higher resolution ($\sim1''$) image, which separates the radio pair better.
The noise in slightly higher in this image.}
\end{deluxetable*}
\clearpage

Six of the SMA sources 
(GOODS~850-7 (GN04), GOODS~850-9 (GN19), GOODS~850-11a (GN12),
GOODS~850-15a and GOODS~850-15b (GN07), GOODS 850-17)
constitute one-half (or, in the case of GOODS~850-15a and 
GOODS~850-15b, a whole) of a radio pair having separations 
between the two radio sources $<5''$. 
Thus, in Table~\ref{tab5}, for each of these pairs
(except GOODS~850-15a and GOODS~850-15b, which are
both already in the table), we give the radio 
measurements of the second radio source below that of the 
first radio source and label it ``comp". We note that
because GOODS~850-17 is such a close radio pair, we decided to
use the Gaussian fit from the higher resolution ($\sim1''$) 
radio image, which separates the radio pair better. The noise 
is slightly higher in this image.

Returning to Table~\ref{tab4}, for the resolved radio sources,
we quote the total radio flux density in Column~8.
Otherwise, we quote the peak radio flux density. 
For the radio sources that are one of a pair, 
we only quote the radio flux density corresponding to the 
SMA source. The radio flux density for GOODS~850-17 is that
measured from the higher resolution image.

We searched a $3''$ radius around each SMA position to find
X-ray counterparts in the 2~Ms 
{\em Chandra\/} catalog of Alexander et al.\ (2003b).
All of the X-ray counterparts are within $1\farcs5$ of the SMA
positions except GOODS~850-17. For GOODS~850-17 the X-ray source 
is at the position of the second radio source in the pair
and not at the position of the SMA source, so we eliminated this match.
We give the soft and hard X-ray fluxes of the remaining matches in
Columns~9 and 10, respectively. 

We searched the literature for spectroscopic redshifts, 
which we summarize in Column~11. We give the origins of the redshifts
in the table notes. Finally, in Column~12, we give the 
millimetric redshifts that we estimate
in Section~\ref{secmilli}.
Where a spectroscopic redshift already exists for the source, we
put the millimetric redshift in parentheses.

\section{Results}
\label{secresults}

\subsection{Multiplicity}
\label{secmult}

The first conclusion that we draw from our SMA observations is that 
the number of bright SMGs in the GOODS-N is significantly overestimated 
in all of the existing SCUBA catalogs. First, as we noted
in Section~\ref{secsample}, there are a number of sources 
in both the W04 and P05 catalogs that are not detected
in the SMA and SCUBA-2 observations. 
Second, there are three SMGs (GOODS~850-11 or GN12, 
GOODS~850-13 or GN21, and GOODS~850-15 or GN07) that are composed of 
multiple fainter objects blended in the single-dish observations 
to form a single, apparently more luminous object.

In Figure~\ref{figfluxes}, we show the fluxes 
from our SMA observations versus the fluxes from the original SCUBA 
observations (W04). For the single SMA sources (red squares), 
the fluxes are in good agreement. For the multiple SMA
sources composing one SCUBA source, 
in each case, the individual SMA sources have lower 
fluxes than the SCUBA source, because the latter's flux is a
blend of all the SMA source fluxes. 
Thus, in Figure~\ref{figfluxes}(a), we show the multiple
sources (blue diamonds)
as separate SMA sources plotted at a common SCUBA 
850~$\mu$m flux. However, in Figure~\ref{figfluxes}(b), 
we plot the multiple sources (blue diamonds) 
after combining their SMA fluxes.  For sources with smaller
separations (i.e., the two double sources), these combined SMA 
fluxes match the SCUBA fluxes well. However, for 
the triple source, where the separations between the sources
approach the SCUBA FWHM, the combined SMA flux 
is larger than the SCUBA flux (not shown; see below), 
which measures only the portion within the beam. 
Wang et al.\ (2011) note that GOODS~850-13a, 13b, 
and 13c actually combine to match two SCUBA sources
(GOODS~850-13 and GOODS~850-23) in the W04 catalog.  We therefore
plot the combined SCUBA fluxes of GOODS~850-13 and GOODS~850-23 
in Figure~\ref{figfluxes}(b). With this adjustment, the total 
SMA flux of the triple source also agrees well with
the SCUBA flux.

\begin{inlinefigure}
\includegraphics[scale=0.5,angle=90]{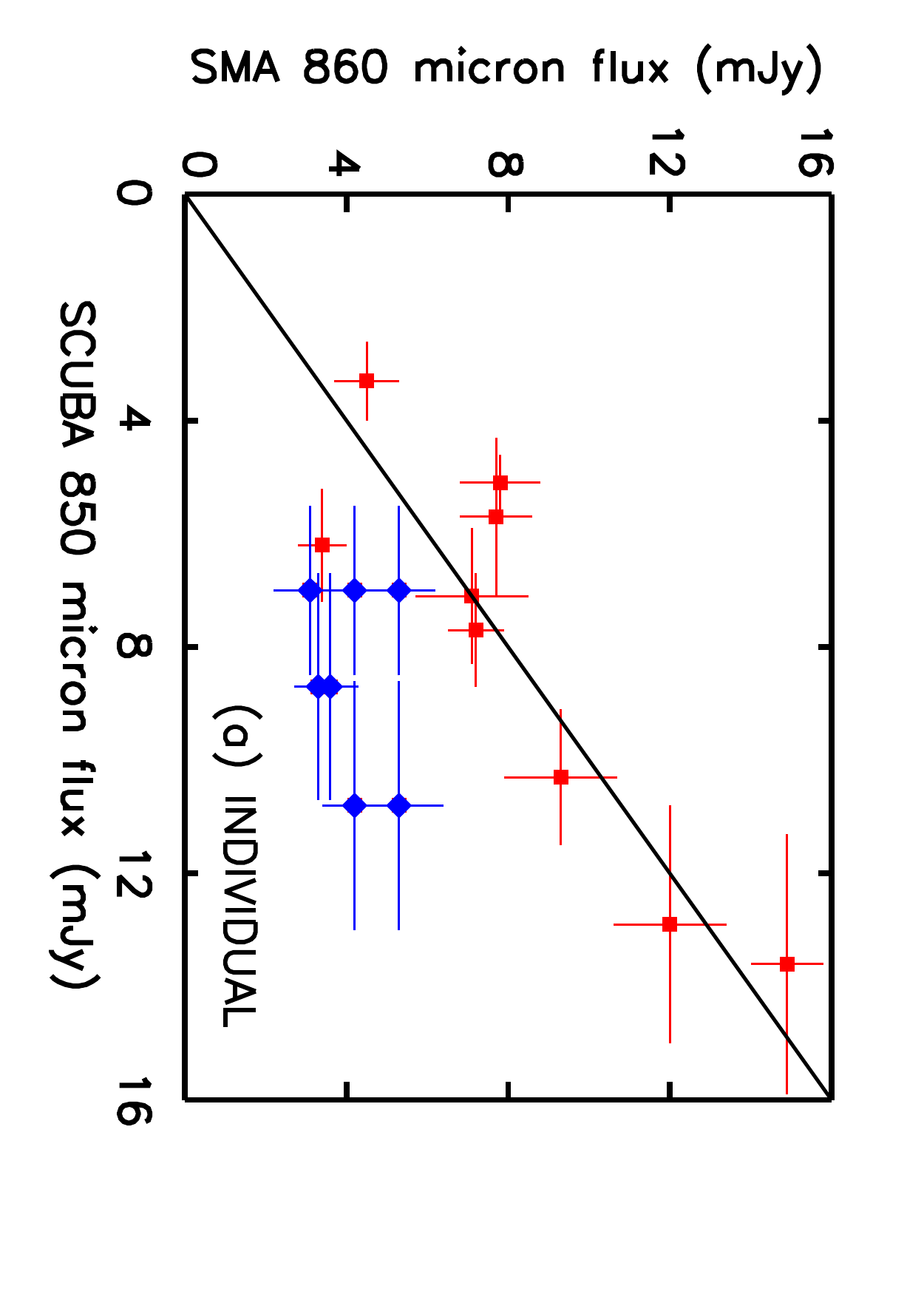}
\includegraphics[scale=0.5,angle=90]{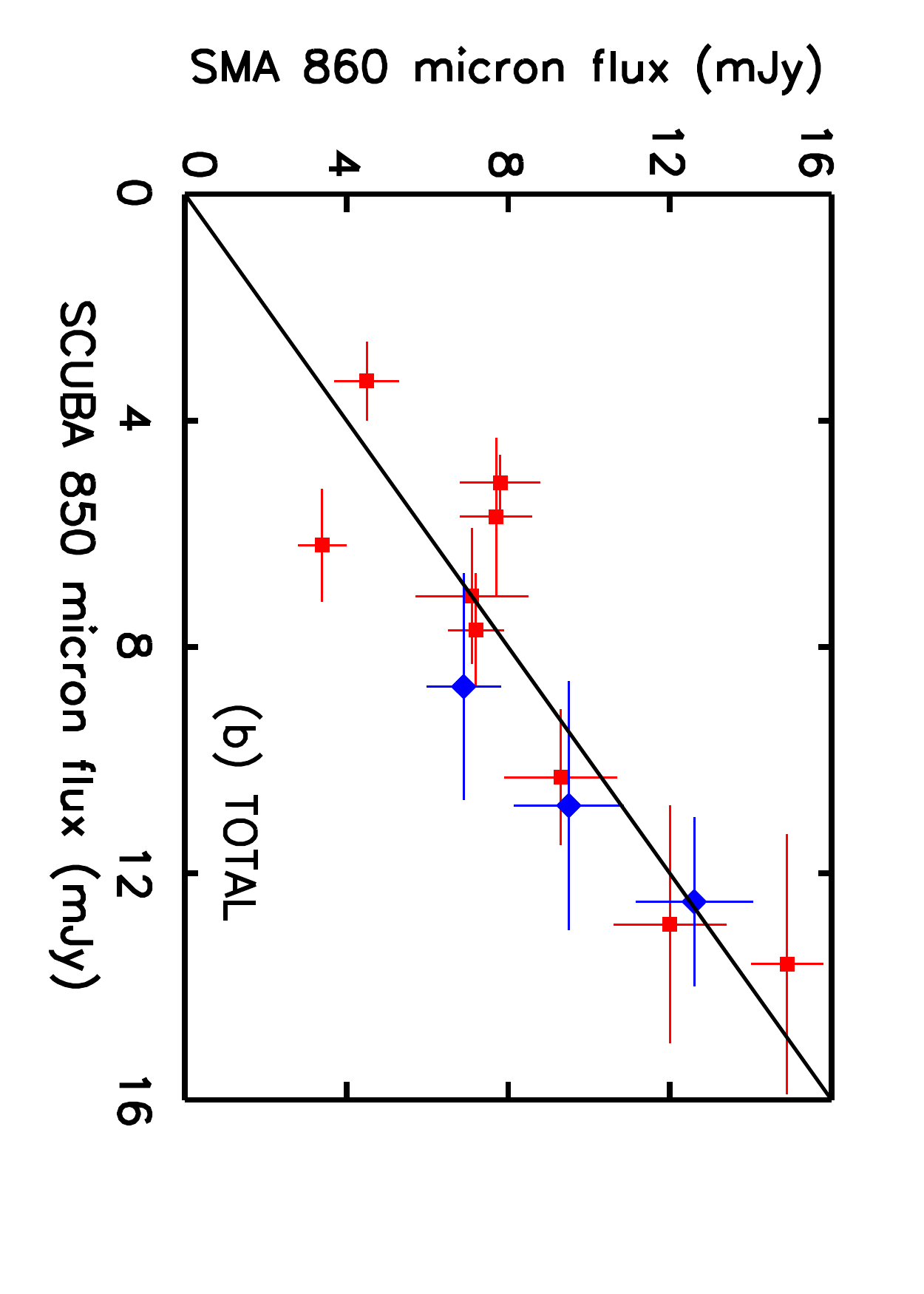}
\caption{
(a) SMA 860~$\mu$m flux vs. SCUBA 850~$\mu$m flux for the SMA
sample in the GOODS-N. SCUBA sources found to be single in the 
SMA images are shown with red squares with $1\sigma$ uncertainties.
SCUBA sources found to be multiples in the SMA images are shown with 
blue diamonds with $1\sigma$ uncertainites. 
In these cases the individual SMA fluxes are lower than 
the SCUBA flux, which is the blended flux of all the individual sources. 
(b) Total SMA 860~$\mu$m flux from the combined individual sources 
vs. SCUBA 850~$\mu$m flux.
The single sources are again shown as red squares, and the multiple 
sources as blue diamonds.  Here we have combined the SCUBA fluxes of 
GOODS~850-13 and GOODS~850-23 (W04) to compare with the combined SMA 
fluxes of GOODS~850-13a, 13b, and 13c (see text for details).
\label{figfluxes}
}
\end{inlinefigure}

For the five sources with 850~$\mu$m SCUBA fluxes above 8~mJy in the 
SMA sample (see Column~4 of Table~\ref{tab3}), the average sensitivity 
of the SMA observations is such that we could differentiate cases where 
two comparable sources were contributing to the SCUBA flux. We find that
three of the five sources are single sources, where the SMA flux 
matches the SCUBA flux, while the remaining two sources are multiples.
Above an 850~$\mu$m flux of 7~mJy, three of the eight sources are 
found to be multiples. These results suggest that 37.5\% of bright 
SMGs are actually blends, though with such small numbers,
the uncertainties in the precise fraction are large (the $\pm1\sigma$ range
is 17\% to 74\%).

Smol\v{c}i\'{c} et al.\ (2012) performed interferometric continuum 
followup with the IRAM PdBI at 1.3~mm of a sample of 28 SMGs
detected in the COSMOS field at 870~$\mu$m with LABOCA. 
Because they did not require
the LABOCA SMGs to be detected at high significance 
($>4\sigma$), and because they had a wavelength mismatch between the 
discovery observations and their followup observations, it is difficult
to make precise comparisons. However, they discovered that 6 of the 
19 LABOCA sources that they detected in their 1.3~mm maps 
broke up into multiple sources. 
[Due to the multiples, they note that, in total, 
this means they detected 26 SMGs, but the
significance of the detections ranges from $4.5\sigma$ (9 sources),
to $4-4.5\sigma$ (7 sources), down to $3-4\sigma$ (10 sources).]
When they added in the 8 LABOCA
sources that had previously been observed and detected 
with millimeter interferometers (CARMA, SMA, PdBI), 
the fraction became 6 multiples 
out of 27 detections (22\%$\pm 9$\%). Given the heterogeneous nature
of the sample, the percentages are quite uncertain, but it is
reassuring that they are broadly consistent with our results.

A comparison with Magnelli et al.\ (2012) illustrates some of the problems that 
can arise from not having high-resolution imaging data.
Magnelli et al.\ (2012) performed a detailed analysis 
of a sample of 61 SMGs (both lensed and unlensed) detected at a range
of significances in a number of fields, including the GOODS-N, at either 
submillimeter or millimeter wavelengths using a variety of ground-based 
telescopes and instruments.
In their Section~3.6, Magnelli et al.\ state that 
GOODS~850-7 (GN04) has two optical counterparts and that the IRAC 
photometry of the two counterparts is consistent with them being at the 
same redshift. They therefore assume that it is an 
interacting pair and derive 
the dust properties of GOODS~850-7 (GN04) by adding 
the MIR, FIR, and radio\footnotemark[2] fluxes of the two ``counterparts'' together.
However, our SMA observations clearly show that the submillimeter emission 
arises from only one of the two IRAC sources (see Figure~A1).

\footnotetext[2]{Although the radio fluxes given in Table~3 of 
Magnelli et al.\ (2012) are said to be 
from the Morrison et al.\ (2010) catalog, 
this does not appear to be the case. For example, for GOODS~850-7 (GN04), 
Magnelli et al.\ list a radio flux density of $89.5\pm6.3~\mu$Jy, but 
the flux density given in the Morrison et al.\ catalog for this source 
is $34~\mu$Jy. 
}

In summary, a uniformly selected, high-significance, and 
well-understood SMG sample with
high-resolution imaging data at the wavelength of detection 
is very important. This becomes particularly essential
when we try to test whether high-redshift sources obey local 
relations, such as the FIR-radio correlation.
Otherwise, the scatter in the relations may easily be overestimated 
and systematic effects introduced.

\subsection{Radio Properties} 
\label{secradio}

As we discussed in the introduction, often SMGs are found to
have multiple candidate radio counterparts. 
However, SMA observations have shown 
that not all of these candidate radio counterparts produce 
submillimeter emission (e.g., Younger et al.\ 2007, 2008, 2009).  

For example, Ivison et al.\ (2007) found that 
the fraction of radio-identified SMGs with multiple candidate
radio counterparts in the SCUBA HAlf Degree Extragalactic Survey 
(SHADES) was $18.5\pm5.3$\% (12/65), of which $15.4\pm4.9$\% (10/65) 
had separations below $6''$.  
However, Hatsukade et al.\ (2010) found 
that their measured SMA flux ($S_{880~\mu{\rm m}}=6.9\pm1.2$~mJy) 
for the SHADES source
SXDF~850.6 ($S_{850~\mu{\rm m}}=8.15\pm2.2$~mJy from SCUBA; 
Coppin et al.\ 2006) came from only one of the three possible 
radio counterparts identified by Ivison et al.\ (2007),
even though all three radio sources have corresponding MIPS 24~$\mu$m
sources (Clements et al.\ 2008). 

\begin{figure*}
\hskip -0.5cm
\centerline{
\includegraphics[width=7.5in,angle=0]{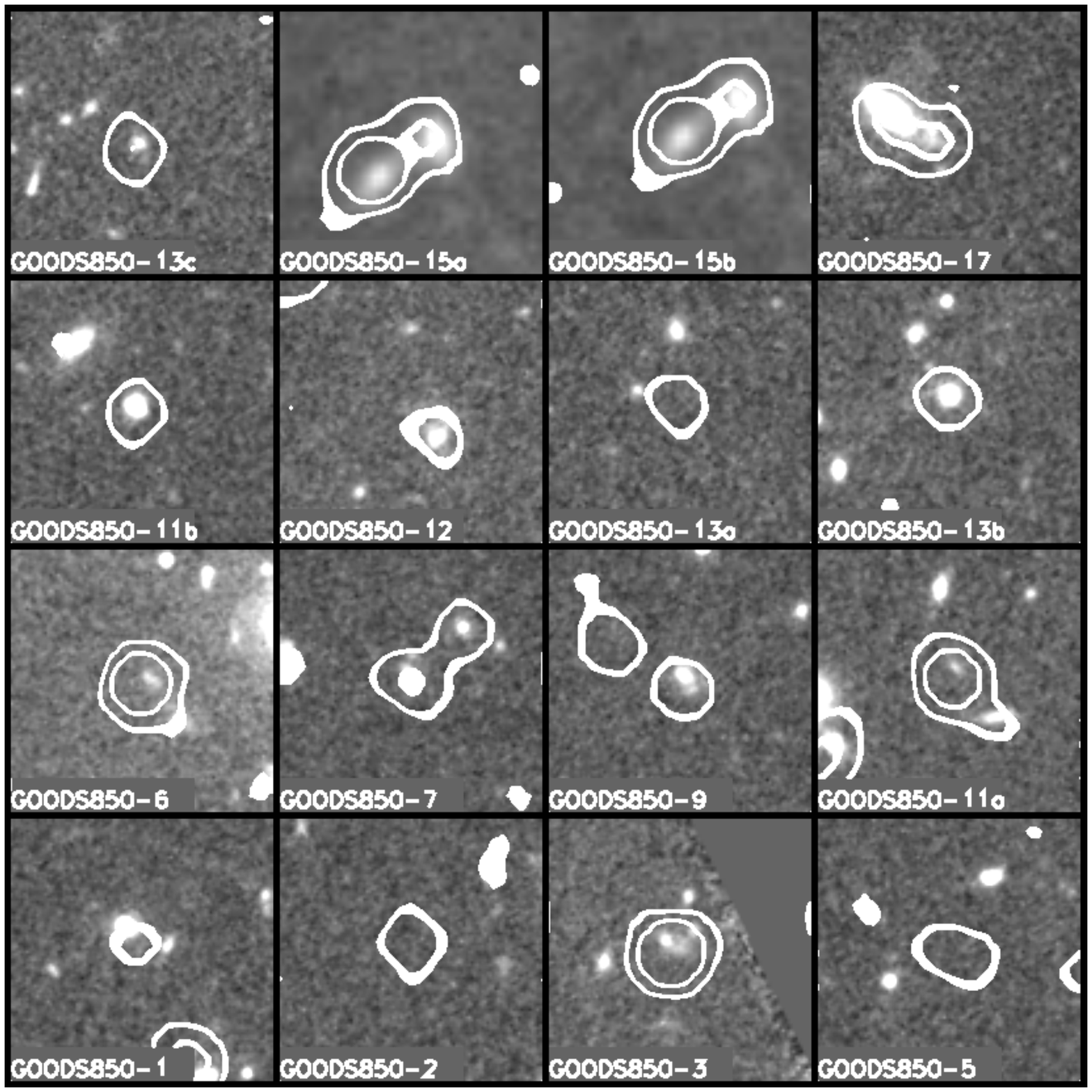}}
\vskip -5cm
\caption{
20~cm contours overlaid on the {\em HST\/} F140W images 
centered on the SMA positions of the SMA sample.
The panels are $10''$ on a side. The F140W images were obtained
from the {\em HST\/} archive and are 811~s exposures. 
The 20~cm image is from Owen13.
GOODS~850-15a, 15b are off the GOODS-N area observed in F140W, so 
for them we show the $K_s$ image from Wang et al.\ (2010). 
The depth of the $K_s$-band image is comparable, but the resolution is poorer.
The lowest contour is chosen to be well above the noise level in the
20~cm image. The second contour is a factor of two higher in surface
brightness to show the position of peaks in the radio. Because
of smoothing, the radio contours are not a good representation of the
20~cm image quality. The properties of the sources that are spatially
extended are given in Table~\ref{tab5}.
\label{radio_thumbs}
}
\end{figure*}

In Figure~\ref{radio_thumbs}, we show 20~cm contours from the 
GOODS-N VLA image of Owen13 overlaid on {\em HST\/} F140W images 
[$K_s$ images for GOODS~850-15a, 15b (GN07), 
where there is no F140W coverage]. 
Each thumbnail in the figure is centered on the SMA position. 
All of the sources in our SMA sample are 
detected in the radio image. 
There are five radio pairs with 
separations between the sources of $<5''$
(GOODS~850-7 (GN04), GOODS~850-9 (GN19), GOODS~850-11a (GN12),
GOODS~850-15a, 15b (GN07), GOODS~850-17).
However, only for GOODS~850-15a, 15b (GN07) does a 
submillimeter counterpart exist for both radio sources in the pair.

In Figure~\ref{radio_sma}, we plot 20~cm flux (Column~8 of 
Table~\ref{tab4}) versus SMA flux (Column~7 of Table~\ref{tab3})
for the sources in the SMA sample. 
There is no obvious dependence of the 20~cm flux on the 860~$\mu$m flux. 
Because all of the SMA sources are detected in the radio at $>5\sigma$,
we can calculate the radio luminosities, subject to the assumptions
about the radio spectral index that we discuss in Section~\ref{highz}, 
once we have the redshifts for the sources.

\begin{inlinefigure}
\centerline{\includegraphics[width=3.8in,angle=90,scale=0.7]{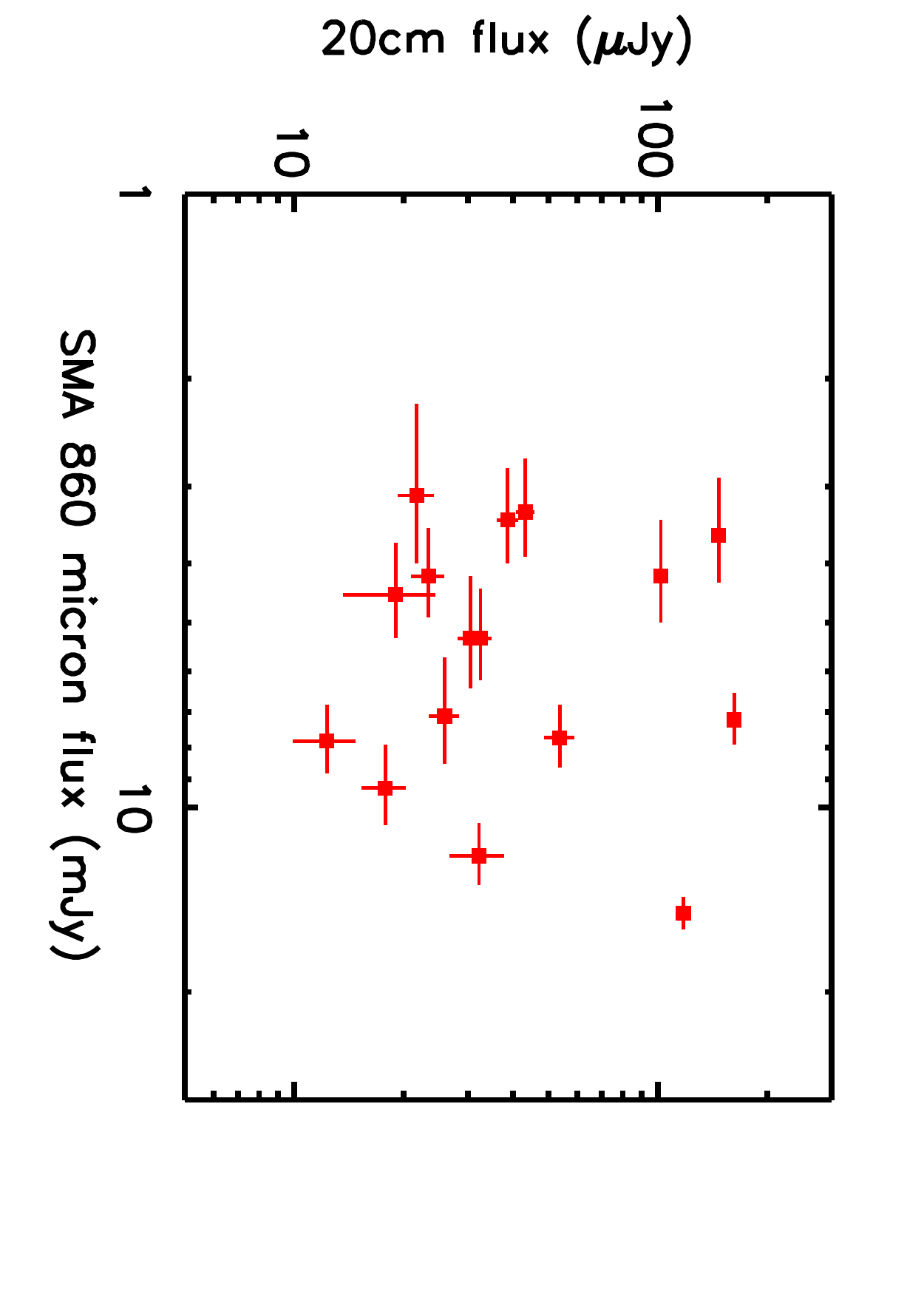}}
\caption{20~cm flux vs. 860~$\mu$m flux with $\pm1\sigma$ uncertainties
for the SMA sample. 
\label{radio_sma}
}
\end{inlinefigure}

\clearpage
\begin{inlinefigure}
\includegraphics[width=3.8in,scale=1.0]{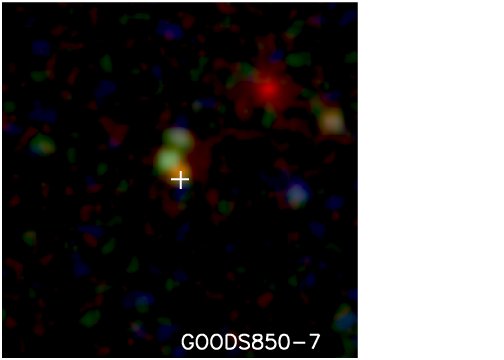}
\vskip 0.1cm
\includegraphics[width=3.8in,scale=1.0]{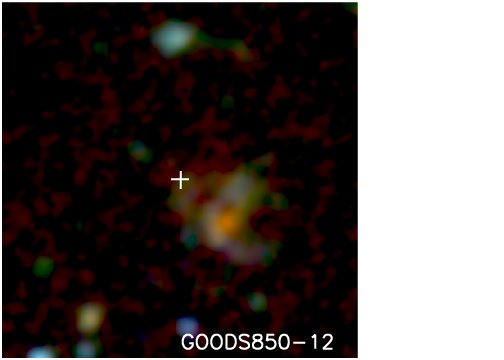}
\vskip 0.1cm
\includegraphics[width=3.8in,scale=1.0]{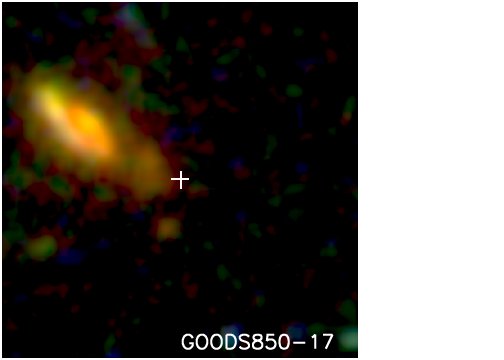}
\caption{
F450W ($B$), F775W ($I$) and F140W ($H$) three-color images
of three SMGs in the SMA sample for which the identifications of the 
counterparts at other wavelengths are not straightforward.
(Top) For GOODS~850-7 (GN04), the counterpart is a chain-like galaxy 
consisting of three components with very varied colors. 
(Center) For GOODS~850-12 (GN15/HDF~850.2),
the submillimeter emission lies near an apparently merging 
system, which is an X-ray source.
(Bottom) For GOODS~850-17, the submillimeter emission lies off the edge 
of a bright galaxy, which is both an X-ray and radio source. 
In this case, we assume the submillimeter emission is not physically 
related to the bright galaxy, and hence we do not assign the redshift 
of the bright galaxy to the SMG.
\label{color_sample}
}
\end{inlinefigure}

\subsection{Redshifts}
\label{secred}

\subsubsection{Spectroscopic}

The accurate submillimeter positions from the SMA allow us to make a more 
critical assessment of the existing spectroscopic redshift identifications. 
We summarize the redshift information from the literature
in Table~\ref{tab4}, where we give both the redshift (Column~11) and the
reference (table notes). 
Unfortunately, even with our accurate positions, there can still 
be uncertanties in assigning the redshifts. 

In Figure~\ref{color_sample}, we show the three most difficult cases from 
our SMA sample using three-color ($BIH$) images marked with white crosses
at the positions of the submillimeter emission.
In the top panel, we show the 
region around GOODS~850-7 (GN04). This source appears
to be a chain galaxy (Cowie et al.\ 1995), 
consisting of three knots with very varied colors. 
The spectroscopic redshift of $z=2.578$ from Chapman et al.\ (2005) 
relates to the upper bluer regions, but the knot corresponding to the 
submillimeter emission is much redder. The magnitudes in the 
table correspond to this redder component. Given the 
configuration, we accept the redshift as applying to the
SMG, but we caution that it could still be a chance projection. 

In the center panel, we show the region around 
GOODS~850-12 (GN15/HDF~850.2). The submillimeter emission arises 
from a location near an X-ray source that has a 
redshift of $z=2.737$ from Barger et al.\ (2008). 
The X-ray source, which appears to be part of a merging system with 
outlying debris, was tentatively identified by P06 as the counterpart 
to the SMG. We feel it is likely that the submillimeter
emission is associated with the merging system and therefore accept
the redshift as applying to the SMG, but again we caution that it
could still be a chance projection. 

In the bottom panel, we show the region around GOODS~850-17. 
Here the submillimeter emission lies off the edge of a much brighter galaxy. 
The bright galaxy is an X-ray and radio 
source with a redshift of $z=1.013$ from Barger et al.\ (2008). 
Although it is possible that this is a merging system, in which case 
the redshift would also apply to the SMG, it is also possible that 
the SMG is a chance projection, possibly amplified by gravitational
lensing due to the larger galaxy. Given its $2.3''$ offset from the 
radio and X-ray source, we do not accept this redshift identification 
for the SMG. (Note that Bothwell et al.\ (2012) also did not detect 
GOODS~850-17 with the IRAM PdBI in CO[2-1] at this redshift.)

With the above assignments for these three difficult cases, we have 
spectroscopic redshifts for 10 of the 16 SMGs (8 optical/IR and 2 CO) 
with values ranging from $z=2-5.2$.

\subsubsection{Millimetric}
\label{secmilli}

As we discussed in Section~\ref{secradio}, with our accurate SMA 
positions, we can unambiguously determine the radio counterparts 
to the SMGs, and with the new ultradeep 20~cm data of Owen13, we 
find counterparts to all of the SMA sources above the $5\sigma$ 
threshold of $12.5~\mu$Jy (see Figure~2 and Table~5).
In combination with the spectroscopic redshifts, this allows us 
to test previous efforts to estimate redshifts for the SMGs
using 20~cm to 860~$\mu$m flux ratios in a precise way that was not
previously possible. 
(Following Barger et al.\ 2000, we refer to such estimates 
as millimetric redshifts.) 

In Figure~\ref{fig5:radrat}, we plot the radio to submillimeter
flux ratios versus $1+z$ using spectroscopic redshifts, where available, 
for the SMGs in the SMA sample (black squares).
For the SMGs in the sample without spectroscopic redshifts, we
plot them at a nominal redshift of $z=0.2$ (blue diamonds).
We distinguish sources that contain AGNs based on their X-ray fluxes 
with red boxes, but these are expected to follow the same relation as 
the non-AGNs, since both star-forming galaxies and radio-quiet AGNs obey the 
same FIR-radio correlation (Condon 1992). We find that the SMGs with
spectroscopic redshifts can be fit by a power law (black line). 
We also show on the plot the Arp~220-based model of Barger et al.\ (2000) 
(blue line) and the M82-based model of Carilli \& Yun (1999) 
(red dashed line). The Barger et al.\ (2000) model agrees reasonably 
well with the power law fit over the observed spectroscopic redshift 
range.  We therefore adopt this relation
(Equation~5 of Barger et al.\ 2000) to measure millimetric redshifts
for the SMGs in our SMA sample.

\begin{inlinefigure}
\hskip -0.8cm
\includegraphics[width=3.6in,angle=90,scale=0.7]{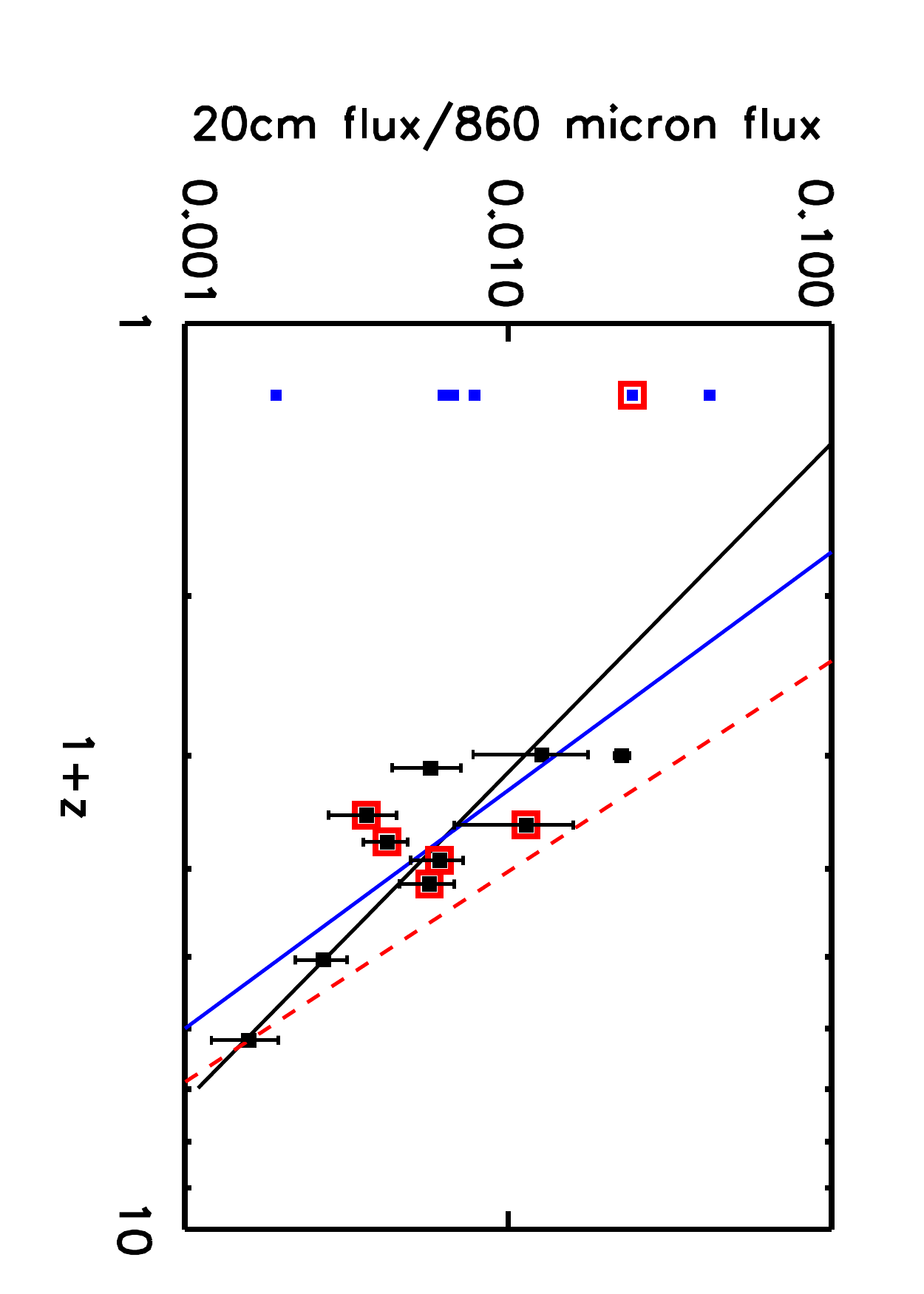}
\caption{20~cm to 860~$\mu$m flux ratio vs. 1+z 
for the SMGs in the SMA sample with spectroscopic redshifts 
(black squares). 
The SMGs in the sample without spectroscopic redshifts 
(blue squares) are shown at a nominal redshift of $z=0.2$. 
SMGs with X-ray detections are marked with red boxes. The
error bars are $\pm1\sigma$. The black line shows a power
law fit to the data. The blue line shows the Arp~220-based model 
of Barger et al.\ (2000). The red dashed line shows the M82-based 
model of Carilli \& Yun (1999).
\label{fig5:radrat}
}
\end{inlinefigure}

In Figure~\ref{fig6:zmilli}, we plot the millimetric redshifts versus 
the spectroscopic redshifts, where available, for the SMGs in the SMA
sample (black squares).
For the SMGs in the sample without spectroscopic redshifts, 
we plot them at their millimetric redshifts in both axes (blue diamonds).
We find that the millimetric 
redshifts generally reproduce the spectroscopic redshifts with a maximum 
multiplicative uncertainty of about 1.4. We show the redshift distributions 
in histogram form at the bottom of the plot. We use the blue histogram to 
show the redshift distribution of the full sample using the combined 
spectroscopic and millimetric redshifts, while
we use the black filled histogram to show the redshift distribution of the
spectroscopically identified sources only.
The spectroscopic redshifts range from $z=2-5.2$,
while the millimetric redshifts of the remaining sources 
range from $z=1.3-4.3$. Within the systematic uncertainties, 
the two redshift ranges could be nearly identical.

\begin{inlinefigure}
\hskip -1.0cm
\includegraphics[width=3.6in,angle=90,scale=0.7]{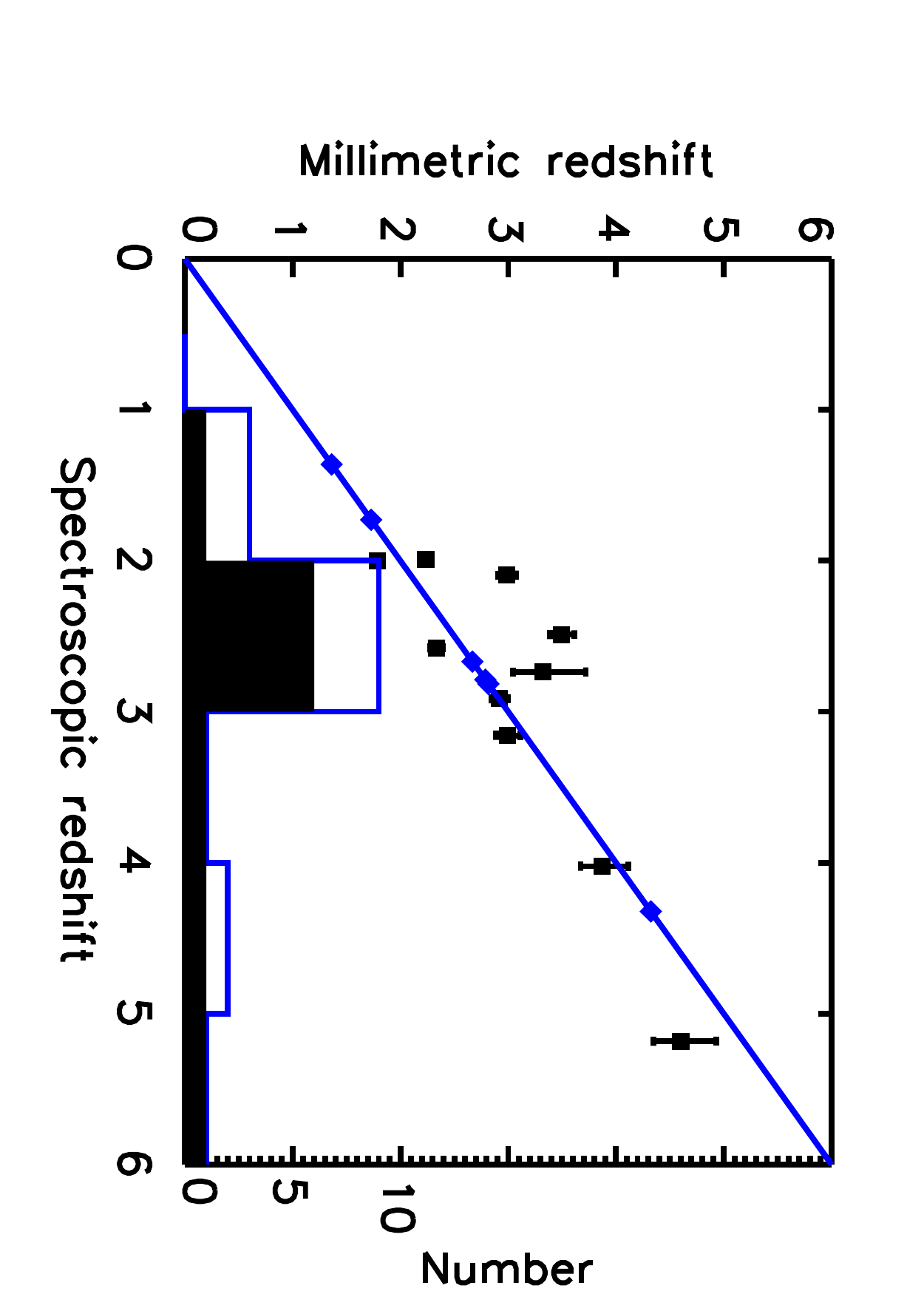}
\caption{Millimetric redshift estimated from the 20~cm to 860~$\mu$m flux 
ratio using the Barger et al.\ (2000) Arp~220-based model
vs. spectroscopic redshift. The SMGs in the SMA sample with spectroscopic
redshifts are denoted by black squares, while those without spectroscopic 
redshifts are denoted by blue diamonds and are plotted at their millimetric 
redshifts in both axes.
The blue histogram shows the redshift distribution of
the full sample using the combined spectroscopic and millimetric
redshifts, while the black filled histogram shows the
redshift distribution of the spectroscopically identified sources only.
\label{fig6:zmilli}
}
\end{inlinefigure}

\begin{inlinefigure}
\hskip -0.8cm
\includegraphics[width=3.6in,angle=90,scale=0.7]{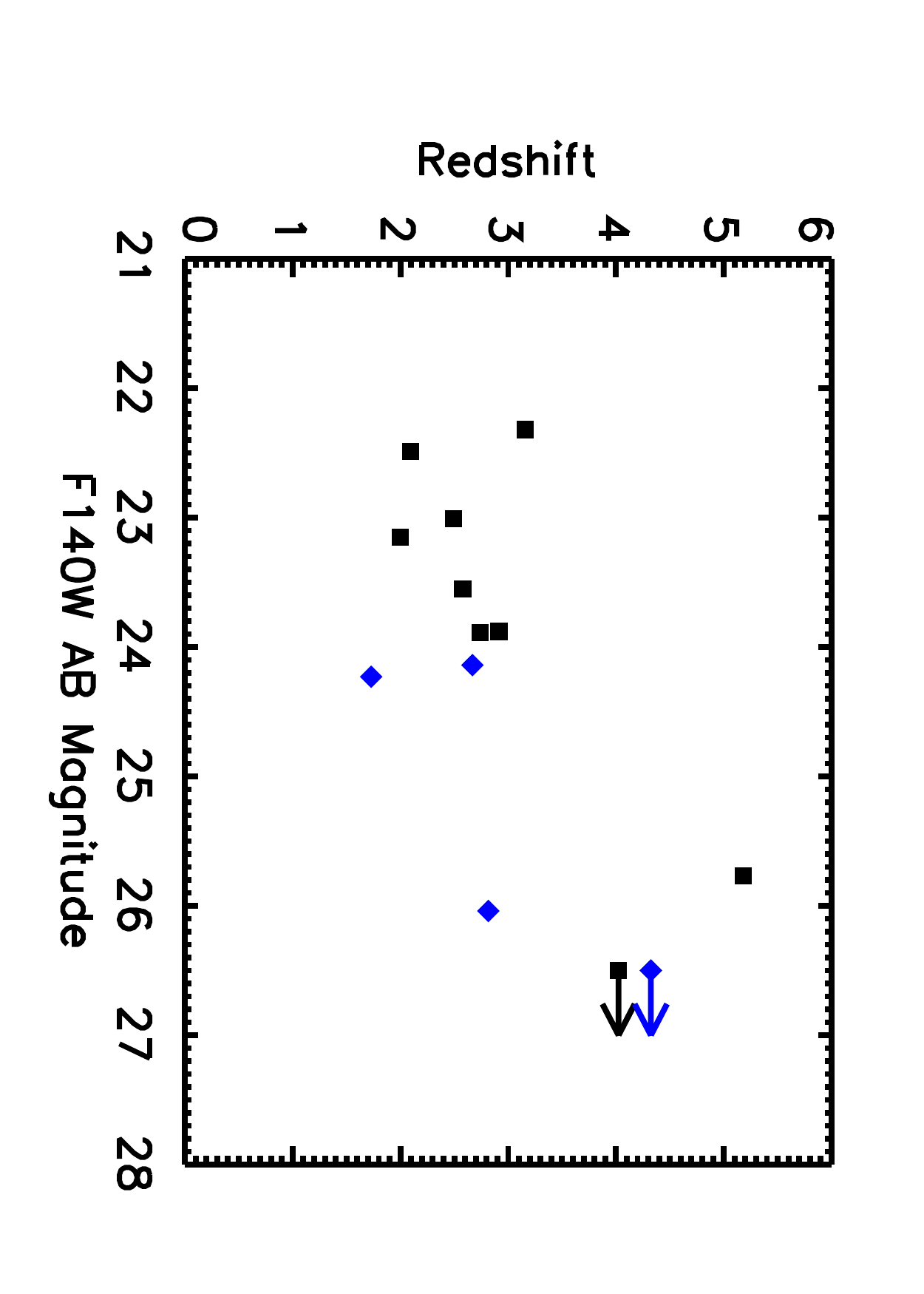}
\caption{Redshift vs. F140W AB magnitude for the SMGs in the 
SMA sample with measured F140W magnitudes. This excludes 
GOODS~850-15a, b (GN07), for which there is no F140W coverage,
and GOODS-17, for which there is too much contamination
from the neighboring source to make a measurement.
The SMGs with spectroscopic redshifts are denoted by black squares, 
while those with only millimetric redshifts are denoted by blue 
diamonds. The right-pointing arrows show the $2\sigma$ 
limit of the F140W image of 26.5.
\label{fig7:nirz}
}
\end{inlinefigure}

\subsection{Optical/NIR Properties}
\label{secnir}

Very high-redshift SMGs are often very faint in the optical/NIR (e.g., 
Wang et al.\ 2009; Cowie et al.\ 2009). Thus, it is worth investigating
with our SMA sample whether there is a particular NIR magnitude limit 
that separates sources at lower redshifts from those at higher redshifts.

In Figure~\ref{fig7:nirz}, we plot redshift versus F140W AB magnitude for
the SMGs in the SMA sample with measured F140W magnitudes.
The black squares denote sources with spectroscopic redshifts, while the
blue diamonds denote sources with only millimetric redshifts.
Two of the SMGs have magnitudes fainter than the $1\sigma$ limit of the 
F140W image of 26.5 (arrows). 
There exists about a 1~mag gap in the plot 
between the lower redshift sources ($z<3.5$) and the mostly 
high-redshift sources, but the total number of sources is very small.

\section{Clean SMA Sample}
\label{secclean}

The primary goal of this work is to understand the properties of a 
highly significant and complete 850~$\mu$m selected sample of SMGs 
for which we know accurate fluxes and positions from the SMA.
We do not require spectroscopic redshifts, but we use them wherever 
available, since they help with determining accurate temperatures 
and spectral energy distributions (SEDs).  
Otherwise, we estimate millimetric redshifts, which we can do 
straightforwardly, since all of the sources in our SMA sample are 
detected in the new ultradeep 20~cm image of Owen13.

Using the measurements in the FIR, submillimeter, and millimeter, 
we now want to compare the shapes of the thermal spectra of our SMGs
with those of local ultraluminous infrared galaxies (ULIRGs). 
In order to do this properly, however, we need to avoid SMGs that 
are composed of multiple sources, each of which may be at a different
redshift, as well as SMGs that are blended with neighboring 
galaxies at FIR wavelengths due to the large beam FWHM sizes of
{\em Herschel\/} ($\sim17''$ at 250~$\mu$m, $\sim 24''$ at 350~$\mu$m,
and $\sim35''$ at 500~$\mu$m; see bottom panels of Figure~A1 in the 
Appendix).
In all of these cases, it would be impossible to disentangle
the different dust temperatures. Finally, the SMGs cannot be
too faint in the FIR to have {\em Herschel\/} detections, or 
else we would not be able to map the peak of the dust spectrum. 

Thus, for this portion of our analysis, we need to remove from 
our SMA sample the multiple sources 
(GOODS~850-11a, b (GN12), GOODS~850-13a, b, c (GN21), 
and GOODS~850-15a, b (GN07)); the sources that are too close to a 
neighboring galaxy for accurate {\em Herschel\/} FIR measurements 
(GOODS~850-2 (GN09) and GOODS~850-17); and the sources that 
do not have {\em Herschel\/} detections (GOODS~850-1 (GN14/HDF850.1) 
and GOODS~850-12 (GN15/HDF850.2)). 
This leaves us with a clean sample of 5 sources, 4 of which have 
spectroscopic redshifts and 1 (GOODS~850-6) of which has only a 
millimetric redshift.
Hereafter, we refer to this as our clean SMA sample.

\subsection{Spectral Energy Distributions}
\label{secsed}

In Table~\ref{tab6}, we list, where available, the MIR through 
millimeter flux measurements for each of the 5 SMGs in our
clean SMA sample, providing the corresponding references for 
the flux measurements in the table notes. 
We show these measurements in Figures~\ref{fig8:sed}(a) and 
\ref{fig8:sed}(b) as different colored symbols (diamonds for the
FIR, submillimeter, and millimeter data, and squares for the MIR 
and radio data) for each source. 
Due to the dense clustering of the data 
points, we only show error bars on the radio data to maintain the 
clarity of the plots. However, the errors on the radio data are
comparable to the symbol size for most of the galaxies.
In Figure~\ref{fig8:sed}(b), we also include the Arp~220 data points
(open triangles) from Klaas et al.\ (1997). From this figure we can
see that the Arp~220 MIR data points drop below
the SMG MIR data points, which are measured at slightly shorter
rest-frame wavelengths. This suggests that the PAH emission
strength is stronger at higher redshifts, as also noted by
Magnelli et al.\ (2012).

\begin{deluxetable*}{cccccccccc}
\renewcommand\baselinestretch{1.0}
\tablecaption{MIR/FIR/Submillimeter/Millimeter Fluxes}
\scriptsize
\tablehead{\multicolumn{2}{c}{Name} & 24~$\mu$m & 100~$\mu$m & 160~$\mu$m & 250~$\mu$m & 350~$\mu$m & 500~$\mu$m & 860~$\mu$m & 1100~$\mu$m \\ 
GOODS & GN/HDF & ($\mu$Jy) & (mJy) & (mJy) & (mJy) & (mJy) & (mJy) & (mJy) & (mJy) \\
\multicolumn{2}{c}{(1)} & (2) & (3) & (4)  & (5) & (6) & (7) & (8) & (9)}
\startdata
850-3\tablenotemark{a} & 06 & $314.5\pm4.1$ & $4.3\pm1.0$ & $25.3\pm2.0$ & $34.2\pm3.1$ & $46.7\pm4.0$ & $27.4\pm4.0$ & $7.2\pm0.7$ & $1.9\pm1.4$ \cr
850-5\tablenotemark{b} & 10 & \nodata & $30.7\pm6.0$ & \nodata & $31.1\pm3.1$ & $42.2\pm4.0$ & $33.7\pm4.0$ & $12.0\pm1.4$ & $5.4\pm1.1$ \cr
850-6\tablenotemark{c} & \nodata & $184.0\pm4.2$ & \nodata & $24.1\pm1.9$ & $54.1\pm3.1$ & $58.2\pm4.0$ & $42.6\pm4.0$ & $14.9\pm0.9$ & $4.1\pm 1.1$ \cr
850-7\tablenotemark{d} & 04 & $322.5\pm4.5$ & \nodata & $12.5\pm1.9$ & $27.3\pm3.1$ & $25.7\pm4.0$ & \nodata & $3.4\pm0.6$ & $2.9\pm1.1$ \cr
850-9\tablenotemark{e} & 19 & $33.9\pm 5.4$ & \nodata & $10.1\pm1.6$ & $21.8\pm3.1$ & $28.0\pm4.0$ & $16.7\pm3.9$ & $7.1\pm1.4$ & \nodata 
\enddata
\label{tab6}
\tablenotetext{a}{The 24~$\mu$m flux is from Magnelli et al.\ (2011).
The separation between the 24~$\mu$m source position and the SMA position
is $0.45''$.
The 100~$\mu$m, 160~$\mu$m, 250~$\mu$m, 350~$\mu$m, and 500~$\mu$m fluxes 
are from Magnelli et al.\ (2012). The 860~$\mu$m flux is from this paper. 
The 1100~$\mu$m measurement is the de-boosted flux from 
Perera et al.\ (2008; AzGN36).}
\tablenotetext{b}{The 100~$\mu$m, 250~$\mu$m, 350~$\mu$m, and 500~$\mu$m 
fluxes were measured by us using the publicly available {\em Herschel\/} 
data. The 860~$\mu$m flux is from this paper. 
The 1100~$\mu$m measurement is the de-boosted flux from 
Perera et al.\ (2008; AzGN03).}
\tablenotetext{c}{The 24~$\mu$m flux is from Magnelli et al.\ (2011).
The separation between the 24~$\mu$m source position and the SMA position
is $0.91''$.
The 160~$\mu$m, 250~$\mu$m, 350~$\mu$m, and 500~$\mu$m fluxes were measured 
by us using the publicly available {\em Herschel\/} data. 
The 860~$\mu$m flux is from this paper. 
The 1100~$\mu$m measurement is the de-boosted flux from 
Perera et al.\ (2008; AzGN05).}
\tablenotetext{d}{The 24~$\mu$m flux is from Magnelli et al.\ (2011).
The separation between the 24~$\mu$m source position and the SMA position
is $0.22''$.
The 160~$\mu$m, 250~$\mu$m, and 350~$\mu$m fluxes are from 
Magnelli et al.\ (2012). 
The 860~$\mu$m flux is from this paper.
The 1100~$\mu$m measurement is the de-boosted flux from 
Perera et al.\ (2008; AzGN16).}
\tablenotetext{e}{The 24~$\mu$m flux is from Magnelli et al.\ (2011).
The separation between the 24~$\mu$m source position and the SMA position
is $0.31''$.  Note that the 24~$\mu$m flux listed in Magnelli et al.\ (2012)
is that of a more distant source (separation $2.96''$), since they did not
have the accurate SMA position.
The 160~$\mu$m, 250~$\mu$m, 350~$\mu$m, and 500~$\mu$m fluxes
are from Magnelli et al.\ (2012). 
The 860~$\mu$m flux is from this paper.}
\label{hermes}
\end{deluxetable*}

At longer wavelengths,
the dust emission from local luminous infrared galaxies is well
described (e.g., see Klaas et al.\ 1997)
by optically thin, single temperature modified blackbodies,
$S_\nu\propto\nu^\beta B_\nu(T)$, where $\beta$ is the emissivity
parameter and is determined empirically. 
More complex fits that include
MIR data are also often used (e.g., Casey 2012 and references
therein), but we have no data between rest-frame wavelengths of
8 and 40~$\mu$m, so we stay with the simpler model.

\begin{inlinefigure}
\centerline{\includegraphics[width=3.6in,angle=90,scale=0.7]{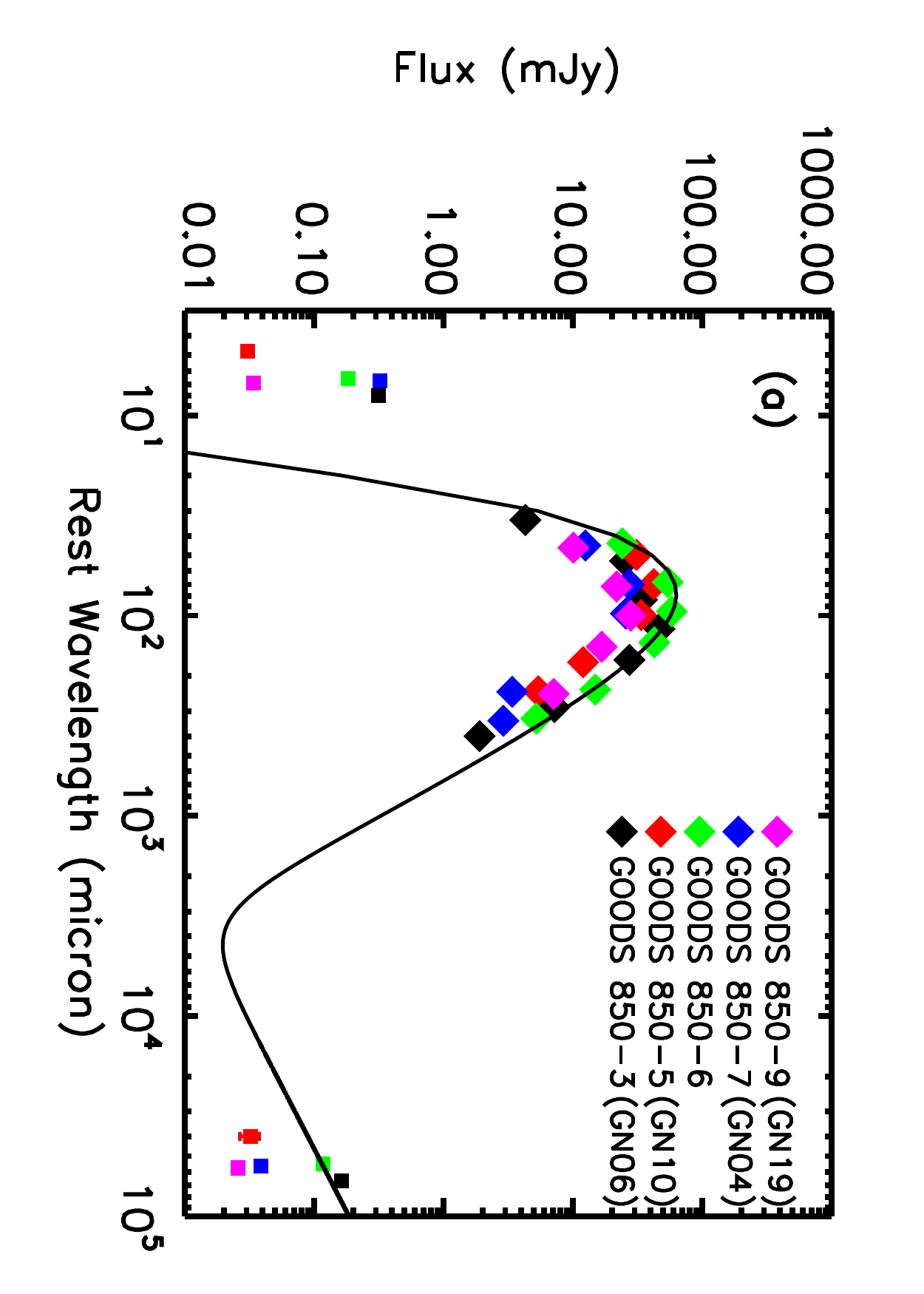}}
\centerline{\includegraphics[width=3.6in,angle=90,scale=0.7]{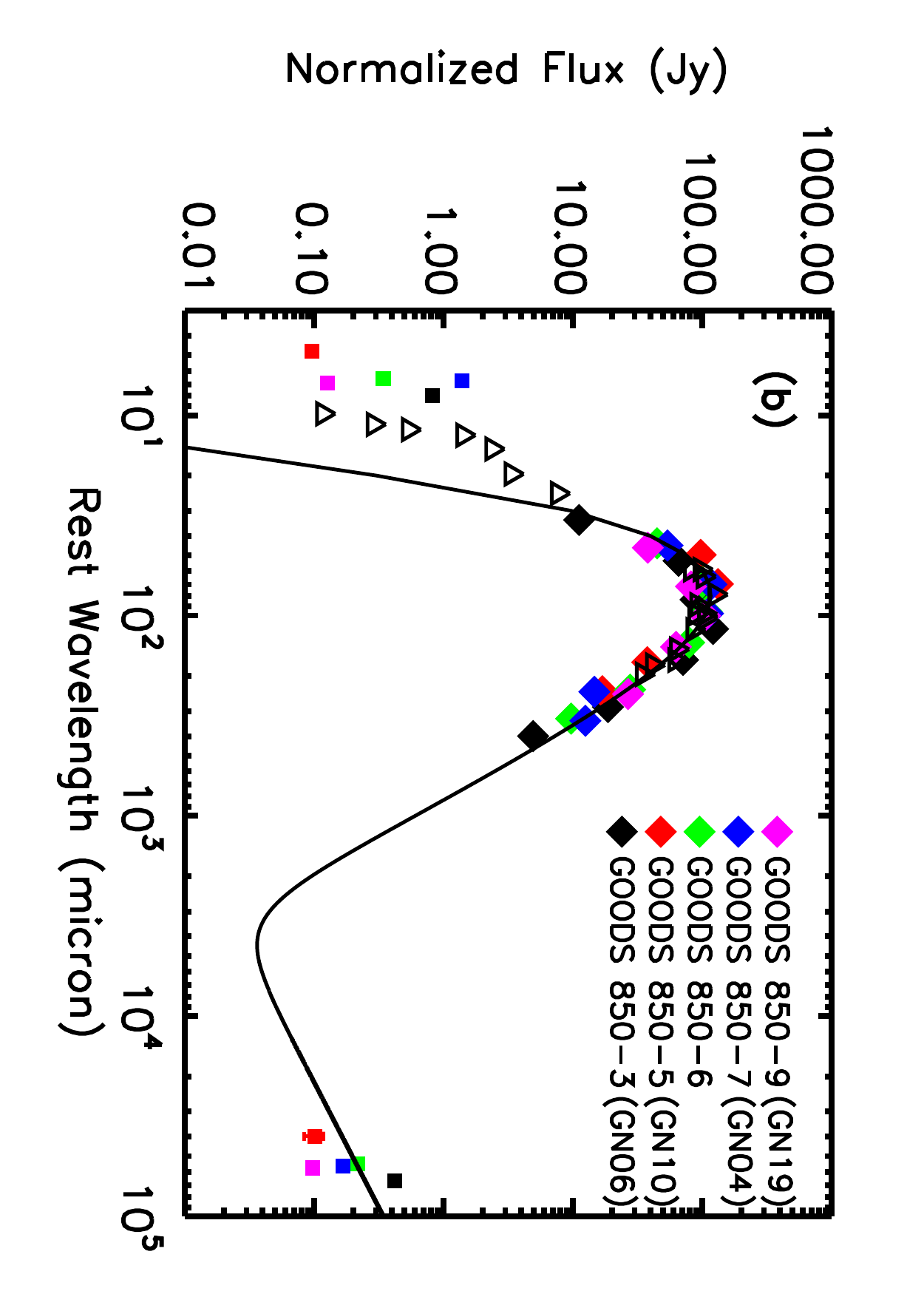}}
\caption{The SEDs for the 5 SMGs in our clean SMA sample (colored diamonds
for the FIR, submillimeter, and millimeter data, and squares for the MIR
and radio data).
For clarity, error bars are only shown for the radio data.
The radio measurements are given in Table~\ref{tab5},
and the MIR through millimeter measurements are given in 
Table~\ref{tab6}.
In each panel the black curve shows a modified blackbody fit to the
Arp~220 data points of Klaas et al.\ (1997) with $\beta=1$ and
a dust temperature of $T_d=47$~K.
(a) Here the modified blackbody is normalized to GOODS~850-6, 
the one source with only a millimetric redshift.
(b) Here the modified blackbody is shown as the original fit to the
Arp~220 data points of Klaas et al.\ (1997; open triangles), and the 
SMGs are normalized to have the same value as Arp~220 at rest-frame
100~$\mu$m, the peak of the distribution where the SEDs are relatively 
flat.
\label{fig8:sed}
}
\end{inlinefigure}

At higher redshifts, Barger et al.\ (2000) used the
Klaas et al.\ (1997) modified blackbody fit to Arp~220 with 
$\beta=1$ and a dust temperature of $T_d=47$~K
to represent their 850~$\mu$m-selected SCUBA sources.
Even for our clean SMA sample sources, which have 
considerably better wavelength coverage than the Barger et al.\ 
sources had, this fit provides a very good 
representation of the data (black curve in Figure~\ref{fig8:sed}). 
Of course, since galaxies do not have single temperatures, such a fit
can, at best, only represent an emission-weighted global average.
In Figure~\ref{fig8:sed}(a), we normalized the 
modified blackbody fit to GOODS~850-6, 
the one source with only a millimetric redshift, which is why 
the radio point for this source lies directly on the synchrotron 
relation. In Figure~\ref{fig8:sed}(b), we normalized the SMGs
to Arp~220 at rest-frame 100~$\mu$m, the peak of the distribution 
where the SEDs are relatively flat. It is clear from both plots 
how similar the distant SEDs are to Arp~220.

\begin{figure*}
\includegraphics[width=3.6in,angle=90,scale=0.7]{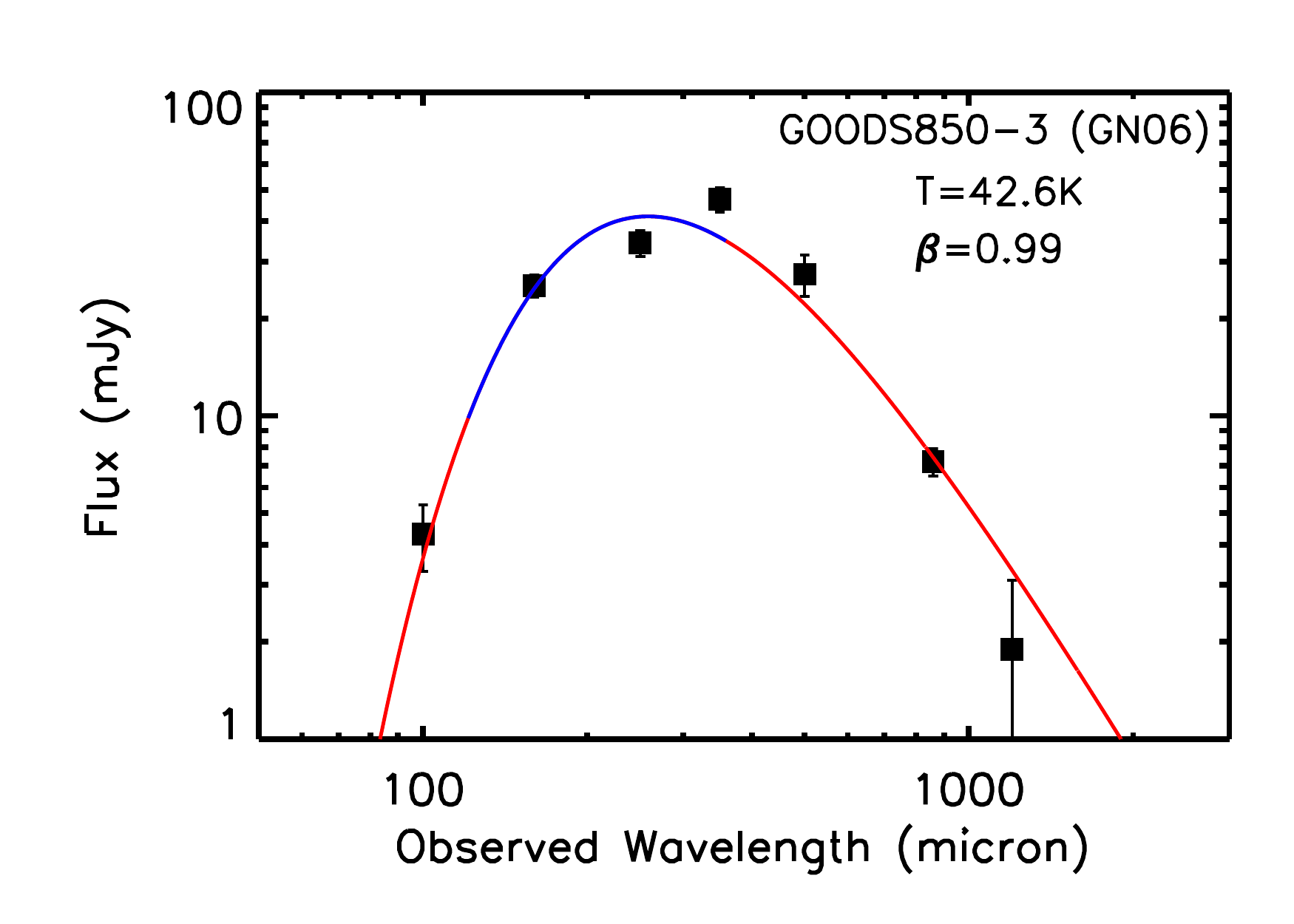}\includegraphics[width=3.6in,angle=90,scale=0.7]{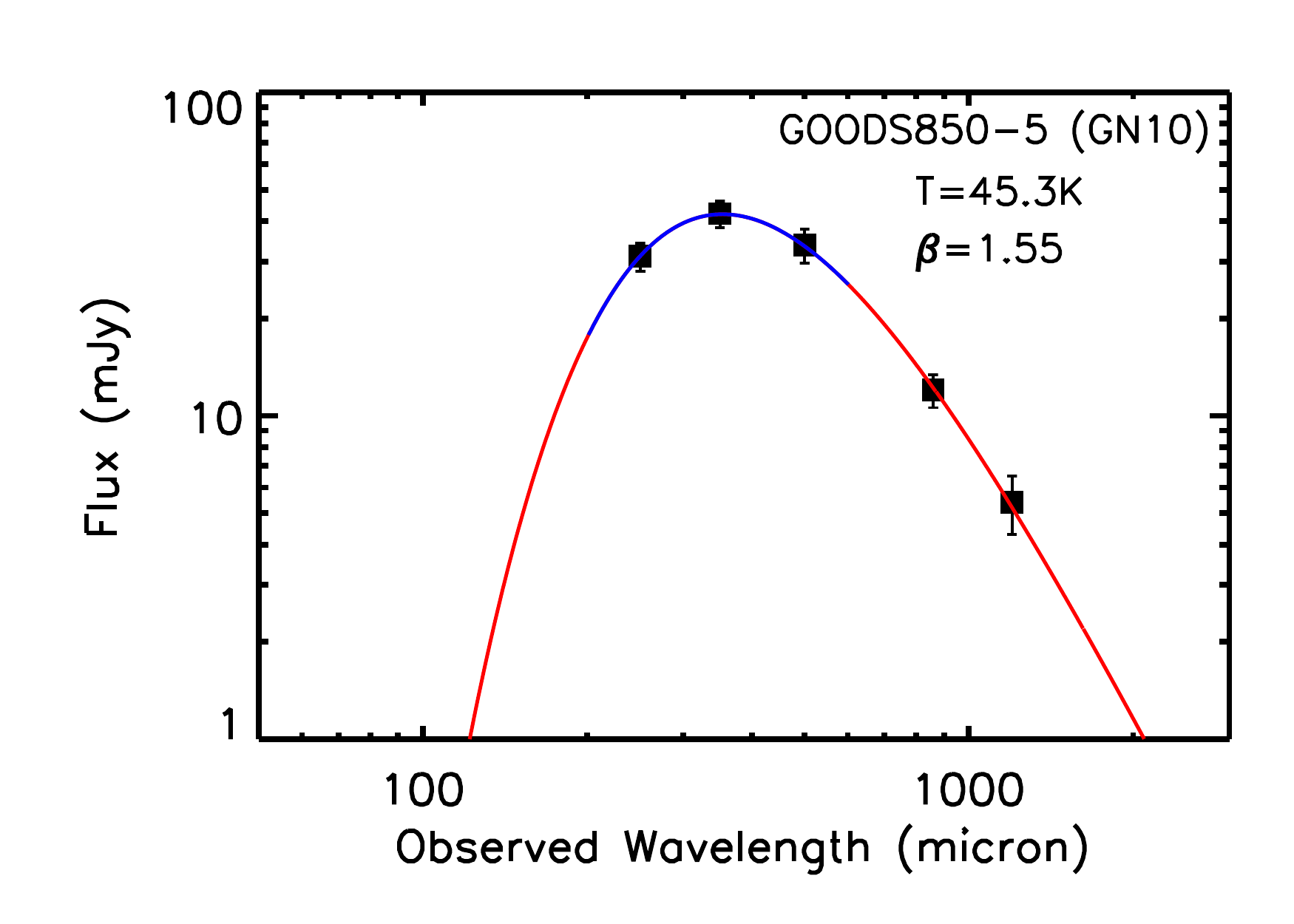}
\includegraphics[width=3.6in,angle=90,scale=0.7]{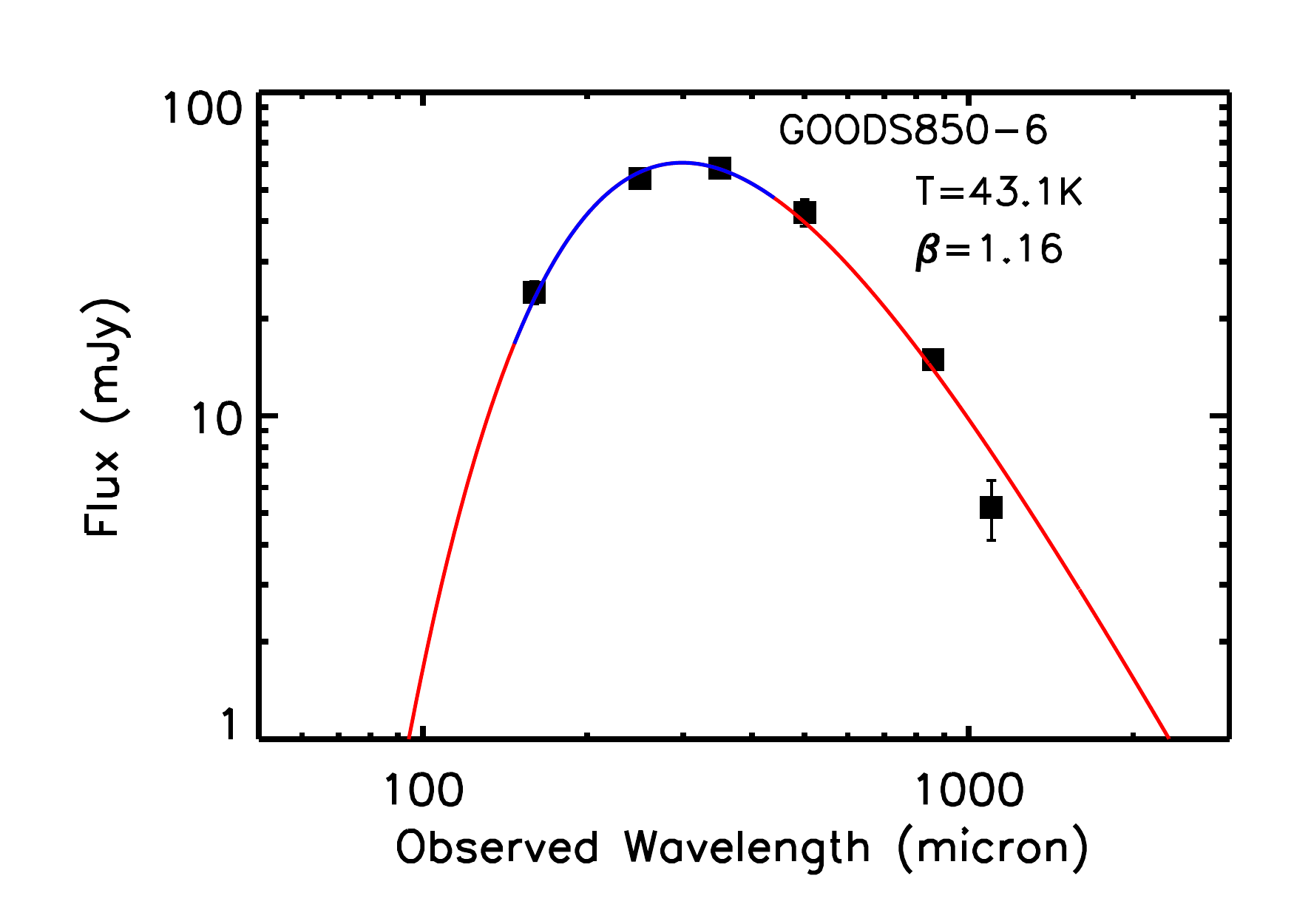}\includegraphics[width=3.6in,angle=90,scale=0.7]{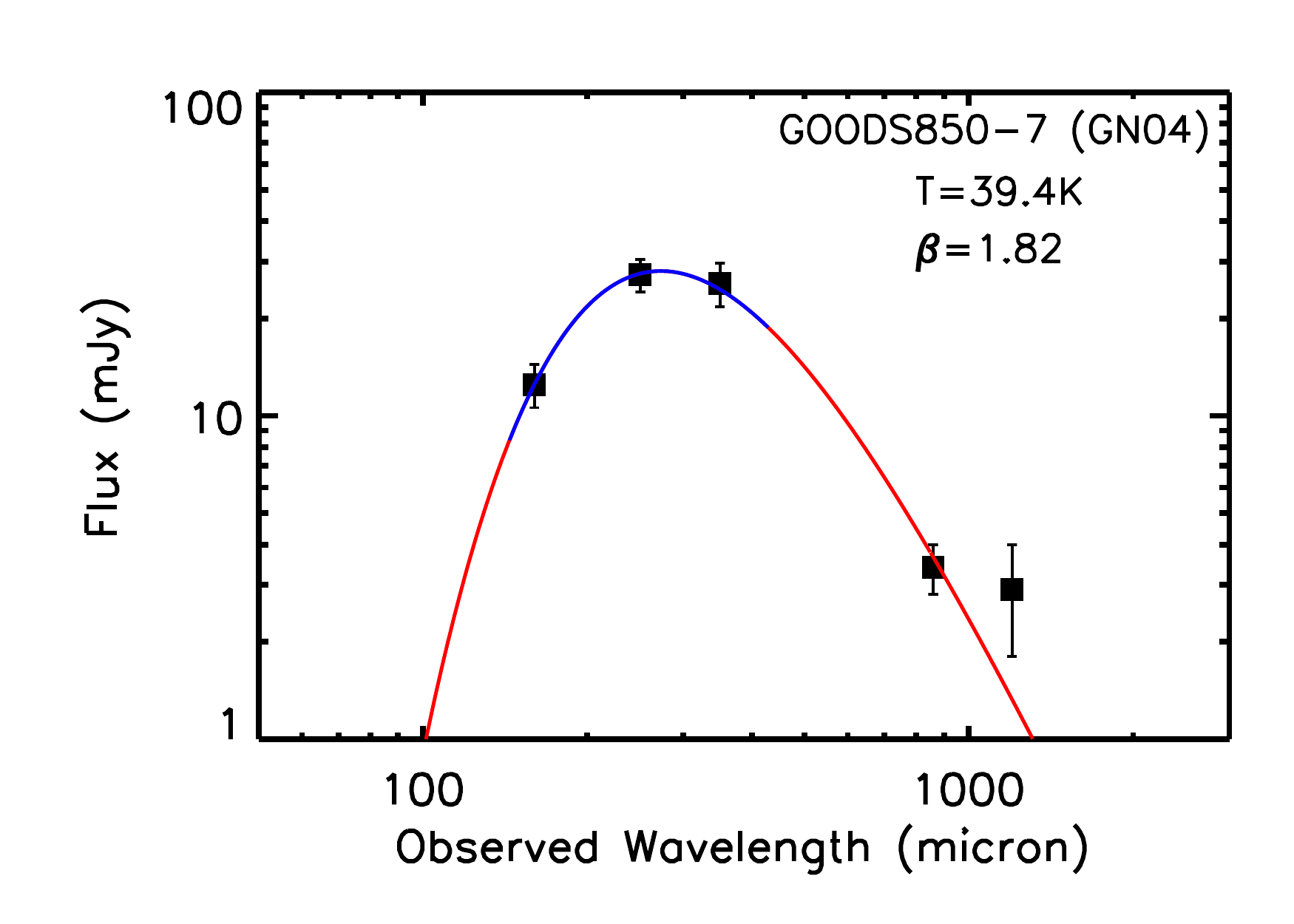}
\centerline{\includegraphics[width=3.6in,angle=90,scale=0.7]{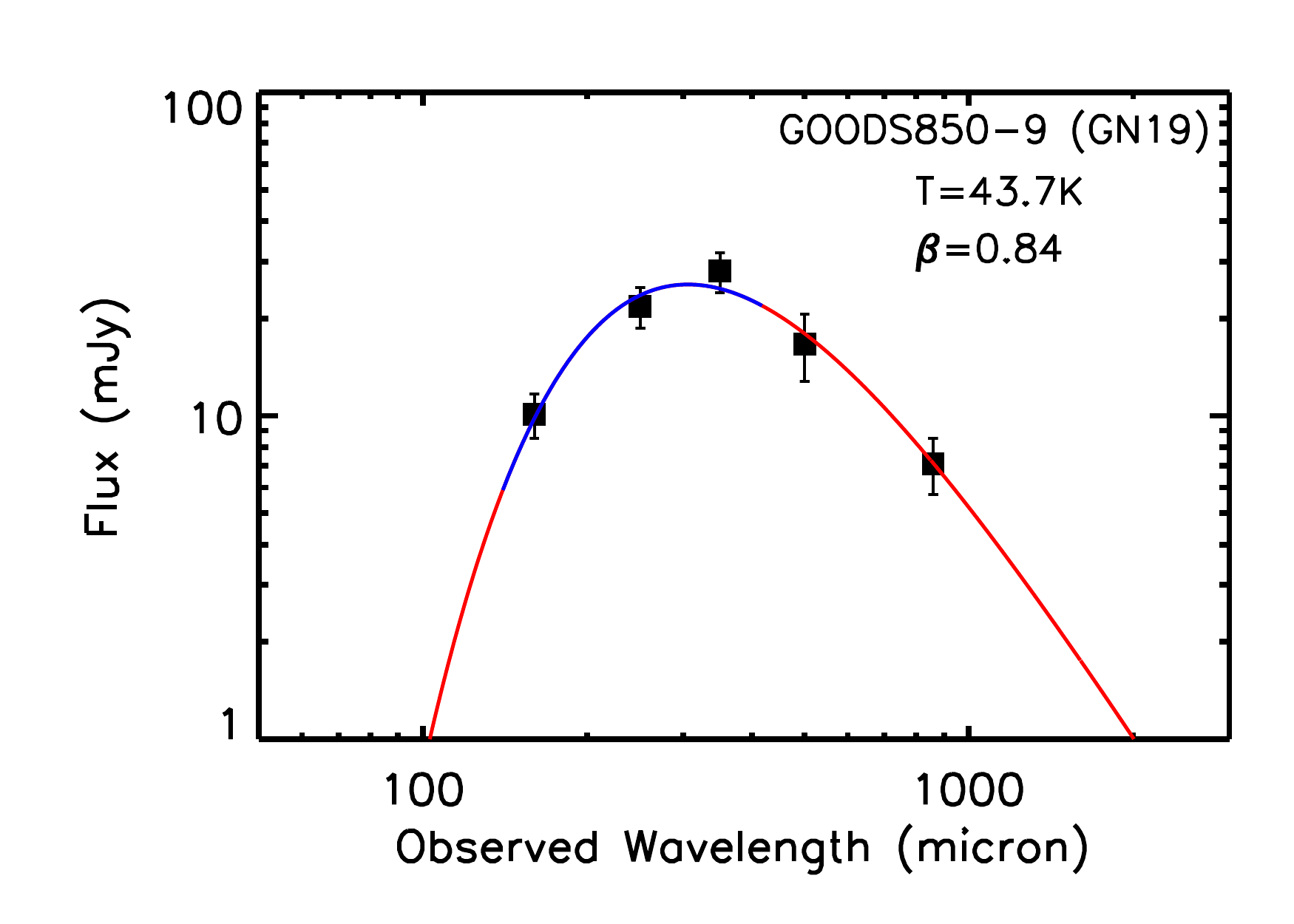}}
\caption{Modified blackbody fits (red curves) to the SEDs
of the 5 SMGs (black squares) in our clean SMA sample,
allowing $\beta$ and $T_d$ to vary. 
The regions corresponding to rest-frame $42.5-122.5~\mu$m are shown in blue 
in each of the panels, since that is the wavelength range of the fits used to 
measure FIR luminosities in Section~\ref{highz}.
The data are taken fom Table~\ref{hermes},
and the error bars are $\pm 1\sigma$. The best-fit values of $\beta$ and $T_d$
are shown in the upper right corner of each panel. 
\label{fig9:fits}
}
\end{figure*}

As noted above, $\beta$ is determined empirically.
Dunne \& Eales (2001) found a constant value of 
$\beta\sim2$ for a sample of local galaxies selected from the
{\em IRAS\/} Bright Galaxy Sample 
that they observed with SCUBA, while
Casey et al.\ (2011) found values ranging from $\beta\sim1-2.5$ 
for a sample of 250~$\mu$m-selected luminous 
galaxies at $z>1$ found with the Balloon-borne Large-Aperture
Submillimeter Telescope (BLAST; Pascale et al.\ 2008).

Assuming $\beta$ to be universal, Magnelli et al.\ (2012) 
performed a global fit using their sample of 61 SMGs by gridding 
the $\beta$ parameter space of $0.1-3.0$ in steps of 0.05.
They then did a $\chi^2$ minimization at each $\beta$ by varying
$T_d$ and the blackbody normalization. They defined the $\chi^2$ 
value at a given $\beta$ to be the sum of the individual 
$\chi_{\rm gal}^2$ values. They applied their global fit to three 
wavelength ranges. They found
$\beta$ values of $0.6\pm0.2$, $1.2\pm0.2$, and $1.7\pm0.3$ for,
respectively, 70~$\mu$m to the
submillimeter, the same, except excluding the 
{\em Herschel\/}-PACS 70~$\mu$m and 100~$\mu$m data, and the same, 
except excluding the 
{\em Herschel\/}-PACS 70~$\mu$m, 100~$\mu$m, and 160~$\mu$m data. 
This observed increase in $\beta$ when excluding shorter wavelength
data led them to conclude that constraints on $\beta$ are very 
sensitive to the wavelength coverage used in the fits, as well 
as to the noise properties of the observations.

Casey (2012) argued that she could constrain $\beta$ 
with $>3$ independent photometric points at rest-frame
$\lambda\ge 200~\mu$m when she used her simultaneous fit 
to a graybody and an MIR power law. When such data are not
available, however, she recommended adopting a 
fixed value of $\beta=1.5$, which is a common approach at
high redshifts (e.g., Chapman et al.\ 2005; Kov{\'a}cs et al.\ 2006; 
Pope et al.\ 2006; Hwang et al.\ 2010).

Here we perform fits to the long-wavelength SEDs of our clean
SMA sample using a modified blackbody and allowing the values of
$\beta$ and $T_d$ to vary. In Figure~\ref{fig9:fits}, we show the 
resulting fits for the sources. We find a mean value of $\beta=1.29$ 
and a mean value of $T_d=42.9$~K for the sample. 
If we instead fix $\beta=1.5$, then we find 
$T_d=36.2$, 46.1, 39.1, 43.1, 37.0~K for GOODS~850-3 (GN05),
GOODS~850~5 (GN10), GOODS~850-6, GOODS~850-7 (GN04), and
GOODS~850-9 (GN19), respectively.
For $\beta=1.0$, we find $T_d=42.6$, 54.7, 45.4, 50.7, and 41.9~K, 
with an average value of 47.1~K. This is nearly identical to the 
Arp~220 value of 47~K when it is fitted in the same way.

\subsection{The High-Redshift FIR-Radio Correlation}
\label{highz}

The FIR-radio correlation is usually parameterized by the 
quantity $q$ (e.g., Helou et al.\ 1985, 1988; Condon et al.\ 1991), 
which is defined as
\begin{equation}
q = \log \left(\frac{L_{\rm FIR}}{3.75\times 10^{12}~{\rm erg~s^{-1}}} \right) - \log \left(\frac{P_{\rm 1.4~GHz}}{\rm erg~s^{-1}~Hz^{-1}} \right) \,,
\end{equation}
where $L_{\rm FIR}$ is the FIR luminosity, and
$P_{\rm 1.4~GHz}$ is the rest-frame 1.4~GHz power.
Here we compute the FIR luminosities over the rest-frame 
wavelength range $42.5-122.5$~{\micron}, which is fully covered 
by the data (i.e., we did the computations using the blue regions 
of the fits in Figure~\ref{fig9:fits}, though we obtained 
identical luminosities when we interpolated the data instead), 
rather than the infrared luminosities over 
the rest-frame wavelength range $8-1000$~{\micron} 
(e.g., Kennicutt 1998; Bell 2003; Ivison et al.\ 2010a, b), 
which would require uncertain extrapolations. 
This also allows us to compare directly with many of the lower
redshift analyses that also use $42.5-122.5$~{\micron}
(e.g., Yun et al. 2001). 
We compute the rest-frame radio power assuming $S_\nu\propto \nu^\alpha$ 
and a radio spectral index of $\alpha=-0.8$ 
(Condon 1992; Ibar et al.\ 2010); that is,
\begin{equation}
P_{1.4~{\rm GHz}}=4\pi {d_L}^2 S_{1.4~{\rm GHz}} 10^{-29}
(1+z)^{\alpha - 1}~{\rm ergs~s^{-1}~Hz^{-1}} \,.
\label{eqradio}
\end{equation}
Here $d_L$ is the luminosity distance (cm) and $S_{\rm 1.4~GHz}$
is the 1.4~GHz flux density ($\mu$Jy). The choice of $\alpha$
may not be appropriate for AGN and also may be problematic
for high-redshift sources. However, we note that, because
we calibrate the relation between the FIR luminosity and the
radio power using this assumption, the final star formation
rate calibrations do not depend on this choice, though
the $q$ values, which represent the intermediate step
in the process, do.

At very high redshifts, this relation must begin to break down,
because the Compton cooling of the relativistic electrons on the
CMB, which increases rapidly
with increasing redshift, will begin to dominate over synchrotron
losses (e.g., Condon 1992). This will decrease the radio power
and increase the value of $q$. The cross-over point occurs
when the energy density in the CMB
becomes comparable to the magnetic field energy
density in the galaxy. For the ULIRGs of the present
sample, where the magnetic field and relativistic energy density
are expected to be extremely high, this may not occur over the observed
redshift range.

\begin{inlinefigure}
\centerline{
\hskip -0.8cm
\includegraphics[width=3.6in,angle=90,scale=0.7]{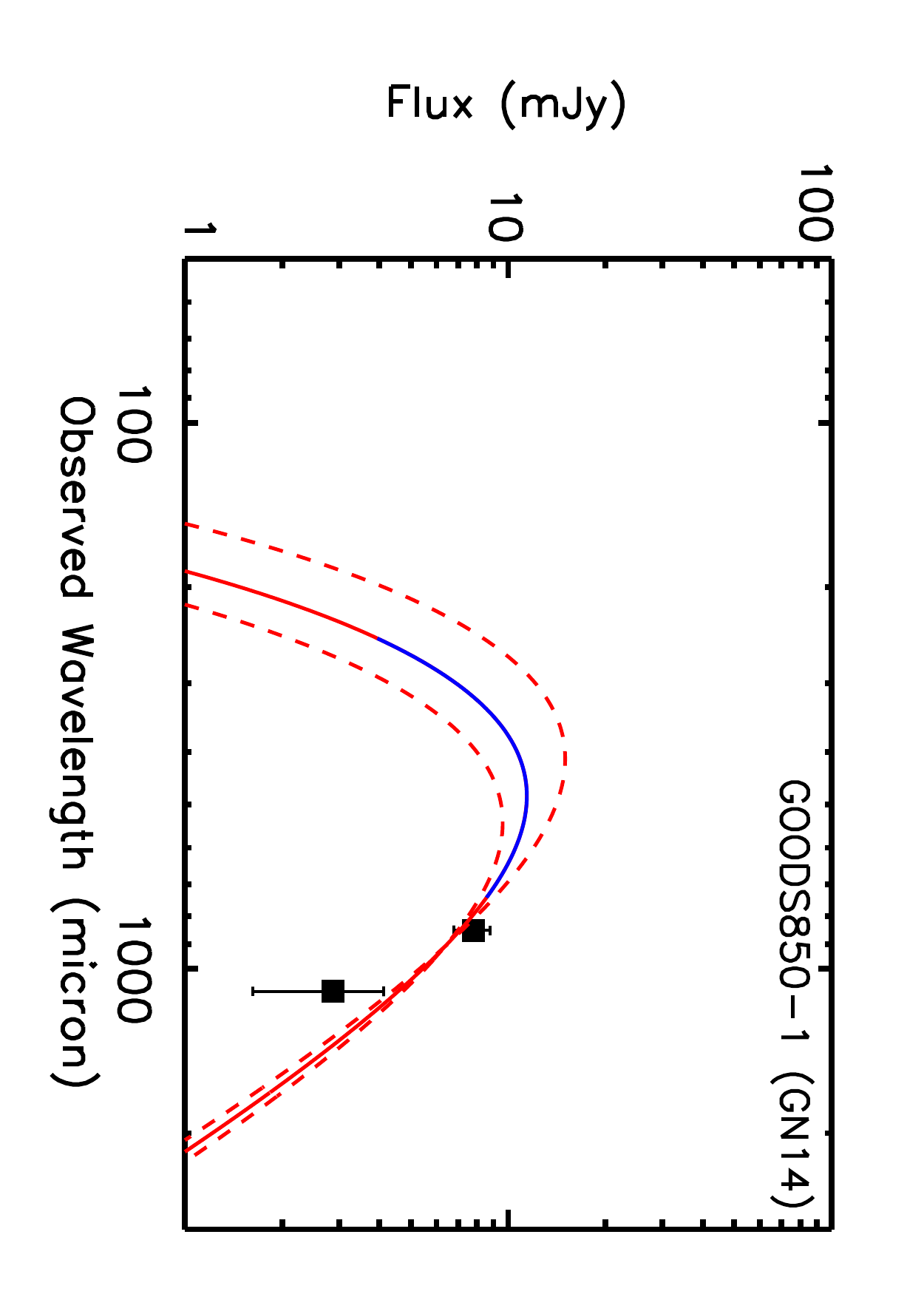}}
\caption{Modified blackbody fit to the 860~{\micron} and 1100~{\micron} data
of GOODS~850-1 (GN14/HDF850.1). The normalization was obtained by 
assuming $\beta=1$ and a temperature of 47~K (red solid curve). 
We also show the fits when the 
temperature is varied from 42~K to 55~K while keeping $\beta=1$ 
(red dashed curves).
The region corresponding to rest-frame $42.5-122.5~\mu$m is shown 
in blue, since that is the wavelength range of the fit used 
to measure the FIR luminosity.
\label{fig10:8501}
}
\end{inlinefigure}

We did not include GOODS~850-1 (GN14/HDF850.1) in our clean
SMA sample, because the {\em Herschel\/} fluxes are too low to be 
confidently measured (see Figure~A1). However, at the known $z=5.183$ redshift
of the source (Walter et al.\ 2012), the observed-frame $860~\mu$m 
lies close to the rest-frame $42.5-122.5~\mu$m band, which means
we can strongly constrain the $q$ value even without the {\em Herschel\/}
data. In Figure~\ref{fig10:8501}, we show fits to our
$860~\mu$m SMA data and the $1100~\mu$m data ($2.87\pm1.25$) 
from Chapin et al.\ (2009; their Table~A3)
assuming $\beta=1$ and $T=47$~K (red solid curve) or
$\beta=1$ and the range $T=42$ to 55~K (red dashed curves). 
The corresponding value of q is 2.34 with a range from 2.22
to 2.50. The calculated q values are not sensitive to the
choice of $\beta$ and the range is primarily determined by
our assumption that the dust temperature is similar to those
at the lower redshifts of our clean SMA sample. 
Using $\beta=1.5$ would change the
q range by approximately 0.01.

\begin{inlinefigure}
\centerline{\includegraphics[width=3.8in,angle=90,scale=0.7]{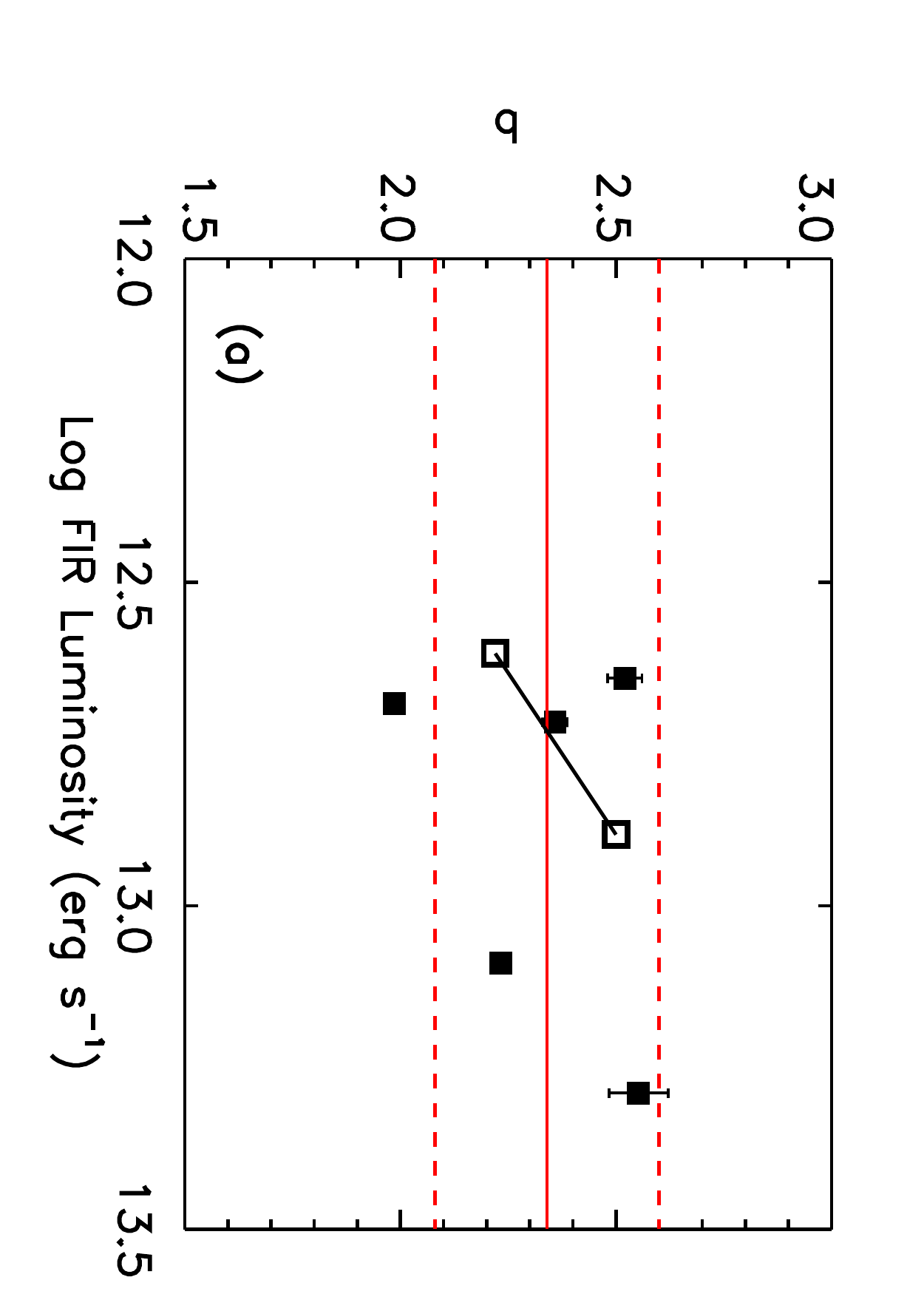}}
\vskip -2cm
\centerline{\includegraphics[width=4.5in,scale=1.0]{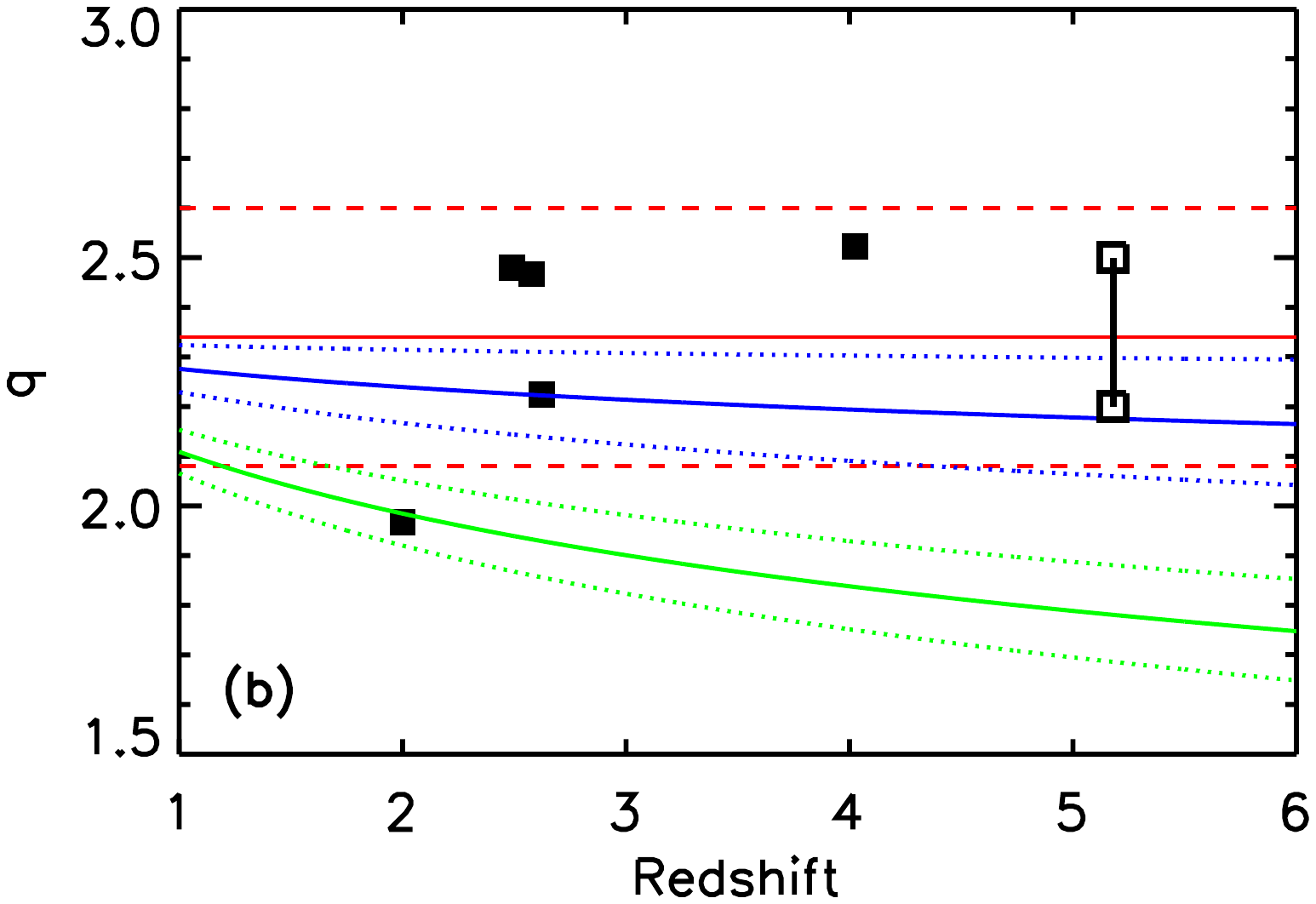}}
\vskip -7cm
\caption{(a) $q$ values and $\pm1\sigma$ error bars for the 5 SMGs 
in our clean SMA sample 
(black squares) vs. their (a) FIR luminosities and (b) redshifts.
The $q$ values correspond to the FIR luminosities computed over
the rest-frame wavelength range $42.5-122.5$~{\micron}, which is 
fully covered by the data. 
The open squares show the range of $q$ values and FIR luminosities
obtained for GOODS~850-1 (GN14/HDF850.1), assuming $\beta=1$ and
a range of temperatures from $T=42-55$~K.
In each panel, the red solid line shows the local 
$\langle q\rangle$ value from Yun et al.\ (2001), 
and the red dashed lines indicate the scatter that 
Yun et al.\ observed in their FIR-radio correlation.
In panel (b), the green solid curve with dotted $\pm1\sigma$ uncertainties
shows the redshift evolution of
$\langle q\rangle\propto (1+z)^{-0.15\pm0.03}$ obtained by
Ivison et al.\ (2010a) from a stacking analysis using {\em Spitzer\/}
and {\em BLAST\/} data in the ECDFS,
and the blue solid curve with dotted $\pm1\sigma$ uncertainties
shows the redshift evolution of 
$\langle q\rangle\propto (1+z)^{-0.04\pm0.03}$ obtained by 
Ivison et al.\ (2010b) from a stacking analysis using
{\em Spitzer\/} and {\em Herschel\/} data in the GOODS-N.
\label{fig11:qcalc}
}
\end{inlinefigure}

In Figure~\ref{fig11:qcalc}, we plot the $q$ values that we calculated 
for the 5 SMGs in our clean SMA sample versus (a) the logarithm of 
their FIR luminosities and (b) their redshifts.
From an {\em IRAS\/} Redshift Survey
sample identified in the NRAO VLA Sky Survey catalog, 
Yun et al.\ (2001) found a local value of
$\langle q\rangle=2.34\pm0.01$ (red solid line) and a scatter in 
the linear FIR-radio correlation of 0.26~dex (red dashed lines).
Our SMGs have $q$ values consistent with this local range, 
with an average value of $\langle q\rangle=2.36$
over the redshift range from $z=2-4.2$.
For an Arp~220 SED, the value of the FIR luminosity calculated
over $8-1000$~$\mu$m is 1.42 times the $42.5-122.5$~$\mu$m
value, giving $\langle q(8-1000)\rangle=2.51\pm0.01$, which
may be compared with the local value of $\langle q\rangle=2.52$
found by Bell (2003).

In Figure~\ref{fig11:qcalc}(a) and (b), we also show the range in $q$ values 
and FIR luminosities (connected open squares) for GOODS~850-1 (GN14/HDF850.1)
that we calculated above assuming $\beta=1$.
We do not use these in our subsequent analysis, but they 
suggest that the invariance in $q$ seen for these 
high-luminosity galaxies extends to beyond $z=5$.

Although many authors have tried to investigate whether the FIR-radio
correlation continues to hold to high redshifts using
infrared data from {\em ISO\/} to {\em SHARC-2} to {\em Spitzer\/} 
to {\em BLAST\/} to {\em Herschel\/} 
(e.g., Garrett 2002; Appleton et al.\ 2004; Kov{\'a}cs et al.\ 2006;
Ibar et al.\ 2008; Garn et al.\ 2009; Sargent et al.\ 2010a, 2010b;
Seymour et al.\ 2009; Ivison et al.\ 2010a, b; Magnelli et al.\ 2012), 
their conclusions have varied.
A major concern for many of these studies is whether radio 
pre-selection may be biasing their results.

Our data extend to higher redshifts than most of these studies 
and are not subject to a radio bias, since all of our sources are 
detected in the radio image. Parameterizing the $q$ values 
for the 5 SMGs in our clean SMA sample 
as evolving as $(1+z)^{\gamma}$ (we normalized to the Yun et al.\ 2001
value locally), we find $\gamma=0.01\pm0.03$, which
is consistent with no evolution. 
For comparison, Ivison et al.\ (2010a) performed a stacking analysis 
using {\em BLAST\/} data
on the Extended {\em Chandra\/} Deep Field South (ECDFS)
at the positions of MIR-selected galaxies from {\em Spitzer\/}
with photometric redshifts
and found $\gamma=-0.15\pm0.03$ (green solid curve with dotted
$\pm1\sigma$ uncertainties) in 
Figure~\ref{fig11:qcalc}(b)).
Likewise, Ivison et al.\ (2010b) performed a stacking analysis
using {\em Herschel\/} data on the GOODS-N 
at the positions of MIR-selected galaxies from {\em Spitzer\/}
spanning $z=0-2$ and matched in infrared luminosity 
($10^{11}-10^{12}~L_\odot$)
and found $\gamma=-0.04\pm0.03$ (blue solid curve with
dotted $\pm1\sigma$ uncertainties).

\section{The Formation History of Extreme Star-forming Galaxies}
\label{secsfr}

The detection in the radio of every source in our SMA sample, together
with a high fraction of spectroscopically identified sources
stretching out to $z=5.18$, as well as millimetric redshifts
for all of the remaining sources, provide us with a sample
where we can measure the
star formation rates (SFRs) per comoving volume to high redshifts 
and compare them with those determined from extinction-corrected 
ultraviolet-selected populations.

If we assume that the FIR-radio correlation is roughly invariant 
for the SMGs with redshifts from $z\approx0-5$, then we can compute 
the SFRs for the individual sources from the 1.4~GHz power.
We caution, however, that we only showed in 
Section~\ref{highz} that this invariance holds at $z=2-4.2$ for the 
more luminous SMGs where we are able to measure the {\em Herschel\/} 
fluxes. Thus, here we have to assume that our results extend down to 
our $860~\mu$m flux threshold of 3~mJy, which is about a factor of 2 
lower than the fluxes of the sources with measured FIR luminosities 
from the {\em Herschel\/} data, and that our results apply out to 
$z=5.18$, where we have less information.


\begin{inlinefigure}
\includegraphics[width=3.6in,angle=90,scale=0.7]{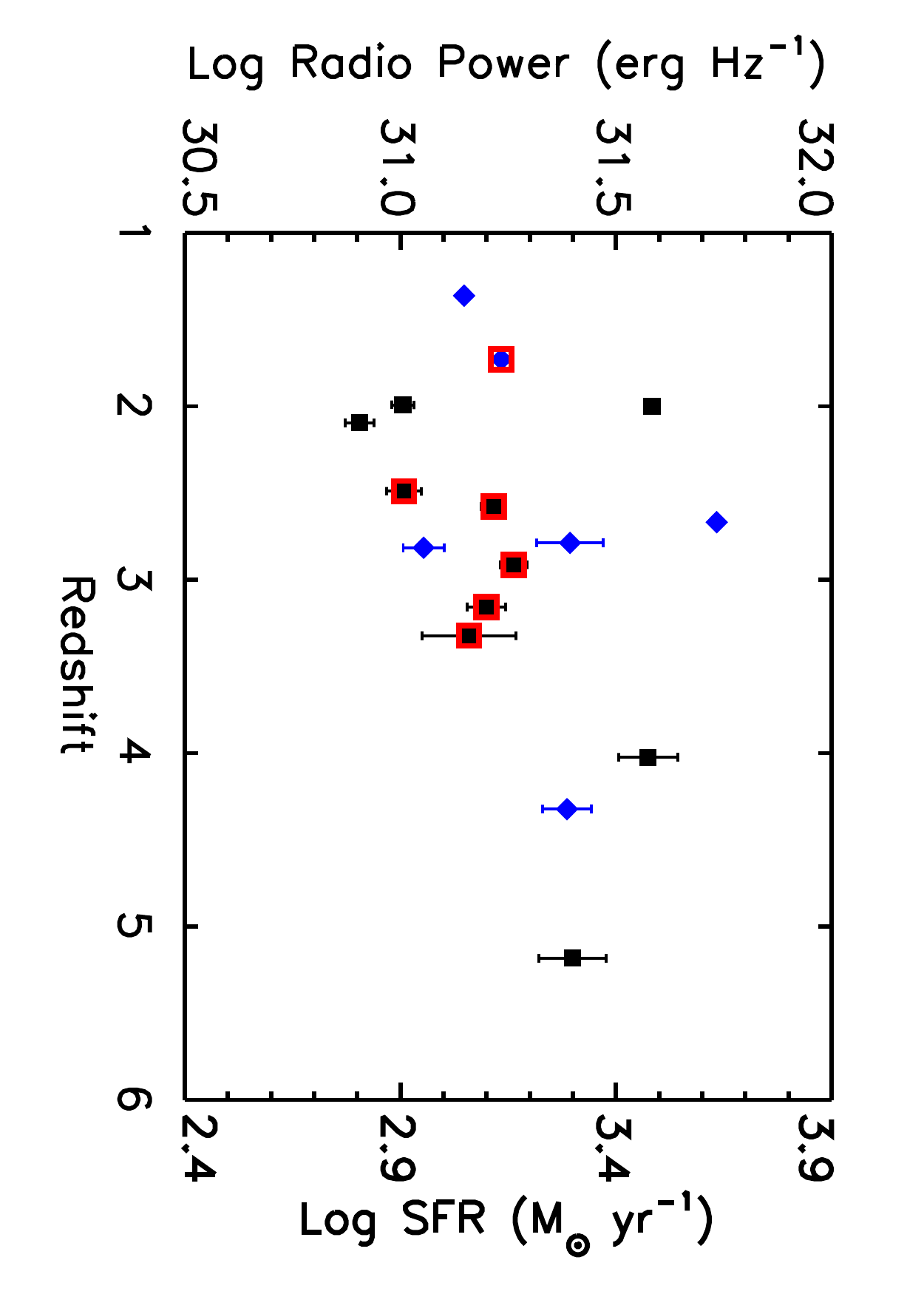}
\caption{Radio power vs. redshift. Sources with spectroscopic redshifts
are shown with black squares. Other sources are shown with blue diamonds
at the millimetric redshift. The right-hand axis shows the corresponding
SFR computed from Equation~\ref{sfr_power} with A$=-28.1$. SMGs with X-ray 
detections are marked with red boxes.
\label{fig12:radpower}
}
\end{inlinefigure}

In Figure~\ref{fig12:radpower}, we show the radio power of the 
SMGs in our SMA sample versus redshift.
We denote sources with spectroscopic redshifts with black 
squares and sources with only millimetric redshifts as blue diamonds.
We convert the radio power to a SFR using the equation,
\begin{equation}
\log {\rm SFR} ({\rm M}_\odot~{\rm yr}^{-1}) = \log {P_{\rm 1.4~GHz}~({\rm ergs~Hz^{-1}})} - A \,.
\label{sfr_power}
\end{equation}
The SFR is for a Salpeter (1955)
initial mass function (IMF) from $0.1-100$~M$_\odot$. 

The normalizing constant $A$ can be calculated from the $q$ value
if we have a relation between the FIR luminosity and the
SFR. Bell (2003) used the Kennicutt (1998) relation between 
the FIR ($8-1000~\mu$m) and the SFR. Applying a small correction 
for the contribution
of old stars to the FIR, they found $A=28.26$. Our results show that
this can be applied to high redshifts. However, this normalization
is about a factor of two lower than Condon (1992) computed based
on the Milky Way. The factor of two range is probably a 
reasonable measure of the systematic uncertainty in the SFR
versus FIR luminosity relation. Following Cowie et al.\ (2012),
we adopt an intermediate normalization of $A=28.1$.

We show the SFRs
on the right-hand axis in Figure~\ref{fig12:radpower}. The values range 
from $700-5000~{\rm M}_\odot$~yr$^{-1}$. Since the galaxies were selected
in the submillimeter, and all are detected in the radio,
the use of the radio power to compute the SFRs should not introduce
any selection biases. Indeed, there are no signs of any strong
dependences of the SFRs on redshift in Figure~\ref{fig12:radpower}.

\begin{inlinefigure}
\vskip -1.25cm
\centerline{\includegraphics[width=4.2in]{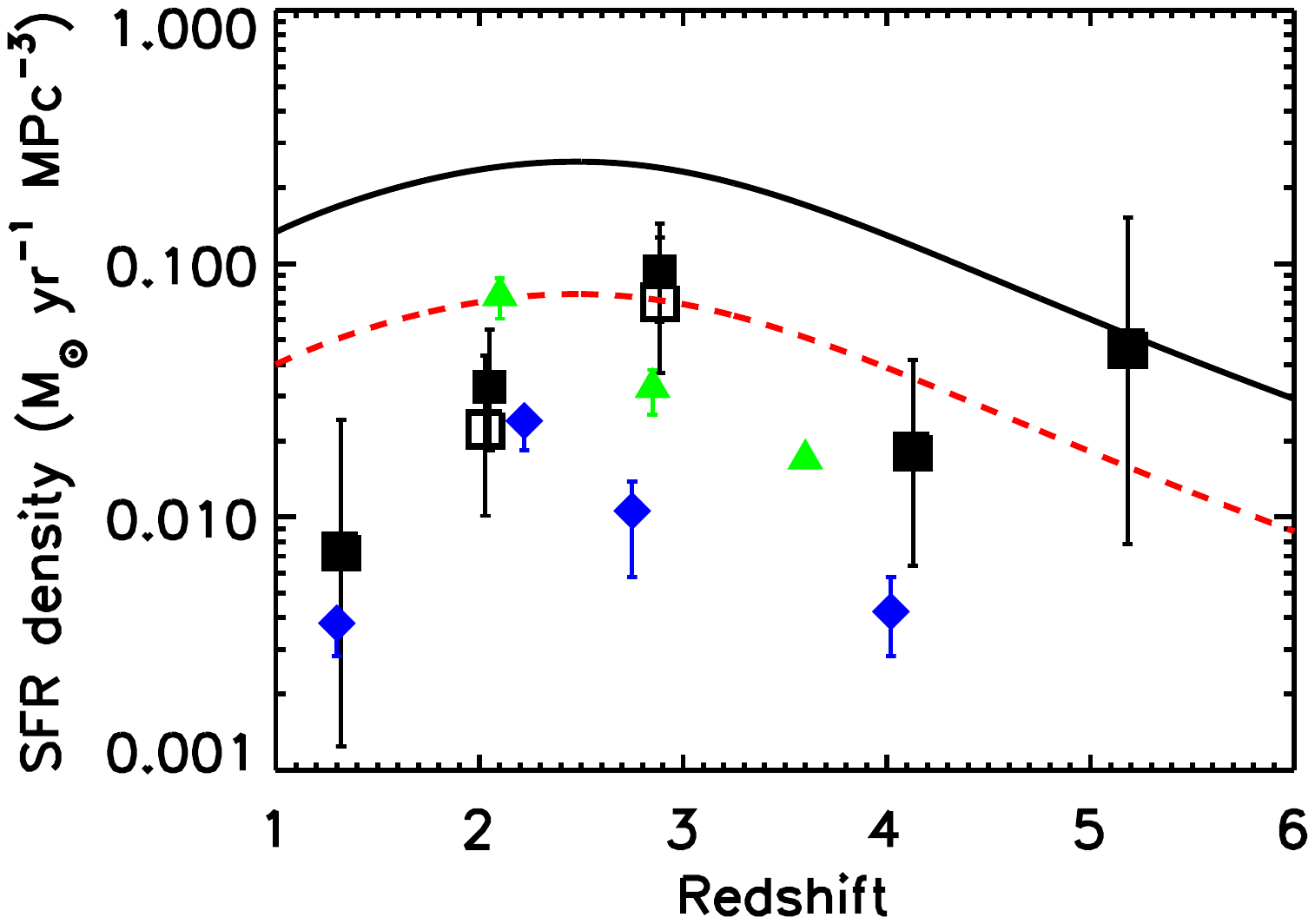}}
\vskip -6.5cm
\caption{SFR per unit comoving volume vs. redshift for our SMA sample 
(black filled squares). The data points are shown at the mean redshift 
of the SMGs in each redshift bin. The error bars are $\pm1\sigma$ 
based on the Poissonian distribution corresponding to the number of 
sources in each bin. These points are not corrected
for completeness and only represent SMGs with 860~$\mu$m fluxes 
above 3~mJy. Thus, they should be considered as lower limits. 
The black open squares show our results after removing five sources
with X-ray detections that could be dominated by AGNs (see text for 
details). The 
black curve shows the SFR density history from Hopkins \& Beacom (2006). 
We scaled their modified Salpeter IMF to our Salpeter IMF.
The red dashed curve shows the Hopkins \& Beacom 
curve renormalized by a factor of 0.3 to match our SMG-inferred value 
in the $z=2-3$ bin. 
The green triangles show the results of 
Chapman et al.\ (2005), and the blue diamonds show the results 
of Wardlow et al.\ (2011), with $1\sigma$ uncertainties.
In both cases, we increased the normalization of their points by a factor of 
1.4 to match our assumed SFR conversion.
\label{fig13:sfr_hist}
}
\end{inlinefigure}

Now we can determine the contributions of the SMGs to the
SFR per unit comoving volume (hereafter, SFR density)
as a function of cosmic time.
The area over which each of the SMGs is detected is
the area over which a SCUBA source 
of that 850~$\mu$m flux could have been
detected at the $4\sigma$ level in the original survey of W04. 
For the SCUBA sources that turned out to be multiple 
SMA sources, the appropriate area for each of the individual 
SMA sources is the area corresponding to the total SCUBA flux
for the composite source. 

We next determined the SFR density in five 
redshift bins of size unity over the redshift range 
$z=0.5$ to 5.5. To do this, we took the individual SFRs in
a particular redshift bin, divided each by the comoving volume 
in that redshift bin calculated using the surveyed area 
for that particular object, and summed them. (Note that the
SMG flux relative to the FIR luminosity is insensitive
to redshift.) We show our results in Figure~\ref{fig13:sfr_hist} 
as black filled squares with error bars that are determined from the 
number of sources in each bin. We note that while the
SFRs of individual sources (see Figure~\ref{fig12:radpower}) 
could be affected by gravitational lensing, the SFR density
would not change, because lensing is surface
brightness conserving, and the surveyed area scales with 
the change in the SFR of the galaxy. However, if lensing were
significant, then the measured SFR density could include the
contributions from more numerous fainter galaxies that lie
below the survey limits.

We compare our results with previous work based on SMG 
samples by Chapman et al.\ (2005, green triangles)
and Wardlow et al.\ (2011, blue diamonds). These authors 
used the same cosmology and Salpeter IMF that we used.
We increased the normalization of their points by a factor 
of 1.4 to match our assumed SFR conversion. 
The Chapman et al.\ SMGs with radio counterparts have 
spectroscopic redshifts for the most part.
They put their SMGs without radio counterparts into the 
$z=3-4$ bin.
The counterparts to the Wardlow et al.\ SMGs come from radio, 
{\em Spitzer\/} 24~$\mu$m, and {\em Spitzer\/} IRAC data and
have mostly photometric redshifts. 
In the $z=2-3$ bin, our results
agree well with those of Chapman et al.
The results of Wardlow et al.\ in this redshift bin 
are slightly low, but this can be partly understood as 
a consequence of the slightly higher flux 
threshold used by Wardlow et al.\ (an $870~\mu$m flux of 4~mJy). 
Wardlow et al.\ also suggest that their results may be low
due to the ECDFS area being underdense (Wei\ss\ et al.\ 2009).
In both cases, the Chapman et al.\ and the 
Wardlow et al.\ results fall below our results at higher 
redshifts. This is easily understood as a consequence of the 
selection bias introduced in their measurements by the radio 
flux limits of their data, which result in the high-redshift sources being omitted.

One of the most difficult issues in deriving the star formation
history is determining the contribution of AGNs to the FIR
luminosity. In computing the points in Figure~\ref{fig13:sfr_hist},
we have assumed that, even when AGN activity is present, it is
not the dominant contributor to the FIR luminosity. This is 
similar to the assumption made by Wardlow et al.\ (2011).
In contrast, Chapman et al.\ (2005) reduced their results by 
30\% to allow for possible AGN contributions. Six of our SMA
sources contain AGNs based on their X-ray luminosities, but in
the case of GOODS~850-12 (GN15/HDF850.2), the radio emission is
extended, suggesting the primary power source may be star
formation (see Table~\ref{tab5}). If we assume the FIR luminosities
of the remaining five sources with X-ray detections are dominated
by the AGN, then that would reduce the SFR per unit comoving
volume in the $z=1-4$ range by a factor of 1.3 (black open squares).
This would also
slightly flatten the star formation history, since it is 
primarily the intermediate redshift bins that are affected.
There may be some extremely Compton-thick AGNs that are not
picked up by the X-ray selection. However, of the 10 sources
that are not X-ray AGNs in the SMA sample, 3 are extended radio
sources (Table~\ref{tab5} lists 3; Momjian et al.\ 2010
has GOODS~850-3 in common with Table~\ref{tab5}),
suggesting that at least in these cases the FIR luminosities
are primarily powered by star formation.

We can now compare our SFR density history with that of
Hopkins \& Beacom (2006; we scaled their modified Salpeter IMF
to our Salpeter IMF), a compilation that is often used 
as a reference (black curve). At the redshifts of interest
($z>2$), the Hopkins \& Beacom results are based on extinction corrected
ultraviolet samples using a common extinction correction. We find that the 
contribution of our SMG sample is about 30\% (red dashed curve) of 
the Hopkins \& Beacom SFR density. Our SFR density is strictly 
a lower limit to the contribution from the overall SMG population, 
since we have not corrected for incompleteness in the original 
sample (both missing sources in the observed flux range and
sources that lie below the 3~mJy flux limit).
However, even the present results show that over the
redshift range $z=1-6$, a large and relatively invariant fraction 
of the overall SFR density is contained in these massively 
star-forming galaxies. It should be emphasized that the present
star-forming galaxies will only be partly included in an 
extinction-corrected ultraviolet 
selected sample, such as that of Hopkins \& Beacom.
Excluding GOODS~850-17, where the neighboring bright galaxy
contaminates the photometry, 7 of the 15 sources have F850LP magnitudes
fainter than 26, 4 of which are not detected at all in the GOODS ACS data.
Thus, roughly half of the star formation seen in this SMA sample would not 
have been included in an ultraviolet star formation estimate.
Thus, determinations of the star formation history from
extinction-corrected ultraviolet selected populations and from 
SMG-selected populations are only partially overlapping and need to be
combined, allowing for the overlap.


\section{Summary}
\label{secsum}

In this paper, we presented our extremely sensitive (sub-mJy rms) 
SMA 860~$\mu$m continuum imaging survey of a complete sample of 
highly significant ($>4\sigma$) SCUBA 850~$\mu$m sources with 
fluxes above 3~mJy that were detected in the W04 survey of the GOODS-N.
Using our SMA observations, as well as SCUBA-2 observations of the field 
obtained by C13, we did not detect 4 of the 
sources in the sample, and we ruled out the original SCUBA fluxes for those
sources at the $4\sigma$ level. Using the SCUBA-2 data, we also did not detect
a similar number of highly significant sources in the P05 SCUBA sample.
It is quite common in the literature to see 
studies that include SMGs detected at significances lower
than $4\sigma$; however, our study illustrates the dangers of 
including such sources and emphasizes the need for caution even
when analyzing highly significant SCUBA samples,
if there are are no confirming submillimeter data available. 

More intriguingly, we found that 3 of the SCUBA sources in the 
sample resolved into multiple, fainter SMGs.
We concluded that the positional accuracy that one obtains from 
interferometric submillimeter or
millimeter observations is absolutely critical for making correct
counterpart identifications and that those identifications made 
at wavebands far from the detection waveband or using low spatial 
resolution data may be highly misleading.

We used new ultradeep 20~cm data of the field obtained by Owen13
with the upgraded VLA to find $>5\sigma$ radio counterparts to 
all of the sources in our SMA sample. We used these data to
estimate millimetric redshifts, though we found 
that the bulk of the sources in our sample (10/16) already 
had spectroscopic redshifts in the literature. 

We constructed SEDs for a clean sample
of 5 SMGs that were isolated (so accurate flux measurements
could be made) and had {\em Herschel\/} data, which provided the
critical measurements at the peak of the thermal dust spectrum.
We found these SEDs to be very similar to that of the
local ULIRG Arp~220. 

Using the SEDs, we measured FIR luminosities for the 5 SMGs
over the wavelength range 42.5-122.5~$\mu$m covered
by the data. We used these FIR luminosities together
with the radio power of the sources to determine $q$, the usual
parameterization of the FIR-radio correlation, for each source. 
We included one additional source at $z=5.183$, whose 860~$\mu$m 
flux is close enough to the peak of the thermal dust spectrum to 
be strongly constraining of $q$, even without {\em Herschel\/} data,
and saw it was also consistent with no evolution in the FIR-radio correlation 
with redshift. Our sample is not subject to the radio bias that has 
plagued many earlier studies on this topic.
We found that our sources had $q$ values consistent with the local
range out to redshifts beyond 5.

With the detection of our entire SMA sample in the radio and the
advantage of having a high fraction of spectroscopic redshifts, 
we were able to measure 
the evolution of the SFR density with redshift for our sample and 
compare it with that determined from ultraviolet-selected populations.
These are the first reliable measurements at high redshifts,
since previous results have been highly biased against such objects
due to the radio selection needed to localize the sources.
We found that the contribution from our sample to the overall
SFR density is a substantial and fairly invariant fraction of 
about 30\% of the Hopkins \& Beacom (2006) extinction-corrected 
ultraviolet selected compilation over the redshift range $z=1-6$. 
We emphasize that determinations of the star formation 
history from extinction-corrected ultraviolet selected populations 
and from SMG-selected populations are only partially overlapping, 
due to the extreme ultraviolet faintness of some of the SMGs.

\acknowledgements

We thank Steve Serjeant for a thoughtful review of the paper.
We thank Y.-W.~Tang and the SMA staff for help in acquiring
the data and M.~A.~Gurwell for help with the data reduction.
We gratefully acknowledge support from
the University of Wisconsin Research Committee with funds 
granted by the Wisconsin Alumni Research Foundation and the 
David and Lucile Packard Foundation (A.~J.~B.),
NSF grant AST-0709356 (L.~L.~C.), and
National Science Council of Taiwan grant
99-2112-M-001-012-MY3 (W.-H.~W.).
This research has made use of data from the HerMES project 
(http://hermes.sussex.ac.uk/). HerMES is a Herschel Key Programme 
utilising Guaranteed Time from the SPIRE instrument team, ESAC 
scientists, and a mission scientist. HerMES are described in 
Oliver et al.\ (2012). The HerMES data were accessed through 
the HeDaM database (http://hedam.oamp.fr) operated by CeSAM and 
hosted by the Laboratoire d'Astrophysique de Marseille.


\vskip -0.3cm
\appendix

We present the SMA and multiwavelength images of the SMA sample in 
Figure~A1. GOODS~850-11 (GN12) and GOODS~850-13 (GN21) were already 
published in Wang et al.\ (2011), and GOODS~850-1 (GN14/HDF850.1) in 
Cowie et al.\ (2009).

\newpage

\renewcommand{\thefigure}{A1}

\begin{figure*}
\centerline{
\includegraphics[width=3.6in]{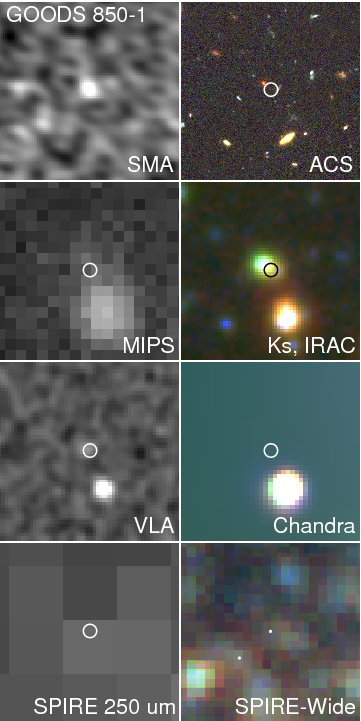}
\includegraphics[width=3.6in]{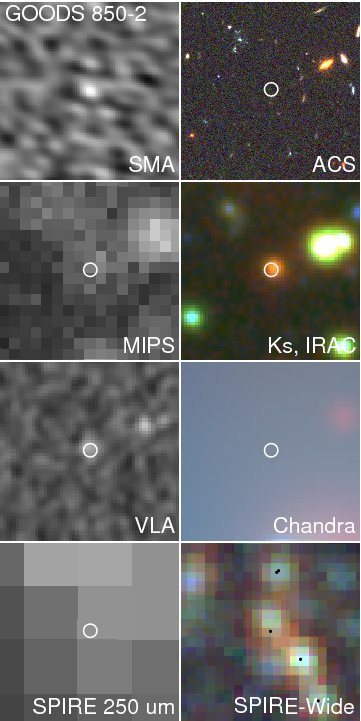}
}
\caption{\small 
{Ultradeep multiwavelength images of the 
SMA sample. The SPIRE-wide panels (bottom rows
right) are $200''$ on a side, while all of the remaining panels are $20''$ 
on a side, except for GOODS~850-11 and GOODS~850-13, where all of the
remaining panels are $25''$ on a side.
The small circles marking the SMA positions have diameters of $1.5''$. 
We show them in either black or white for clarity.
For each source, we show the SMA 860~$\mu$m image (top rows left);
a false-color optical panel made with
{\em HST\/} ACS F435W (blue), F606W (green), and F775W+F850LP (red) 
images (top rows right); the MIPS 24~$\mu$m image (second rows left);
a false-color infrared image made 
with CFHT $K_S$ (blue; Wang et al.\ 2010), IRAC 3.6+4.5~$\mu$m (green),
and IRAC 5.8+8.0~$\mu$m (red) images (second rows right);
the VLA 20~cm image from Owen13 (third rows left);
a false-color X-ray image made with adaptively smoothed 
{\em Chandra\/} $4-8$~keV (blue), $2-8$~keV (green), and 
$0.5-2$~keV (red) images (Alexander et al.\ 2003b) (third rows right);
the SPIRE $250~\mu$m image (bottom rows left); and a wide-field ($200''$) 
color image of the SPIRE 250, 350, and 500~$\mu$m data from the HerMES 
survey of Oliver et al.\ (2012) (bottom rows right).
All of the SMA sources contained in each wide-field
SPIRE image are marked with dots.}
\label{color_thumbs}
}
\end{figure*}

\begin{figure*}[tbh]
\centerline{
\includegraphics[width=3.6in]{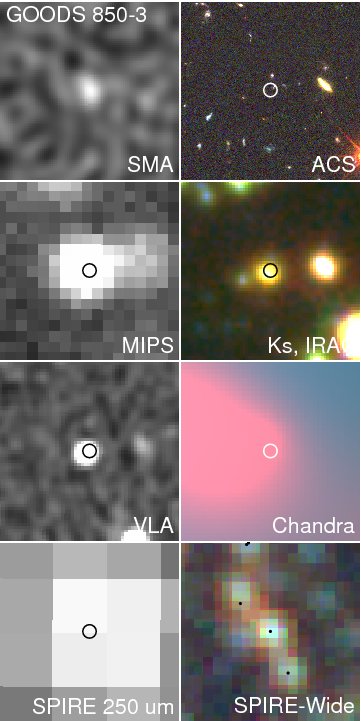}
\includegraphics[width=3.6in]{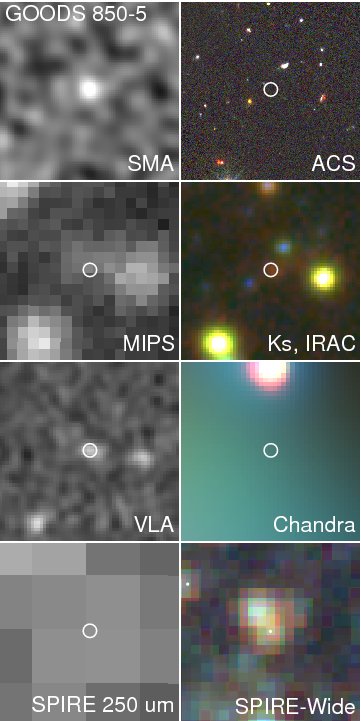}
}
\caption{CONT.}
\end{figure*}

\begin{figure*}[tbh]
\centerline{
\includegraphics[width=3.6in]{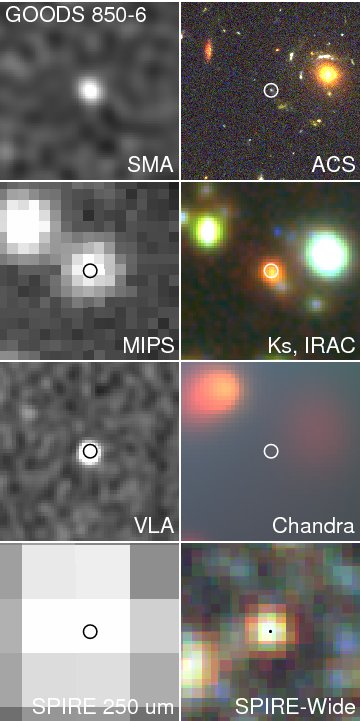}
\includegraphics[width=3.6in]{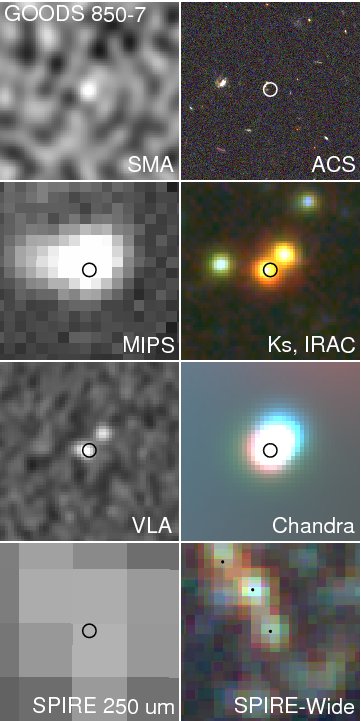}
}
\caption{CONT.}
\end{figure*}

\newpage
\begin{figure*}[tbh]
\centerline{
\includegraphics[width=3.6in]{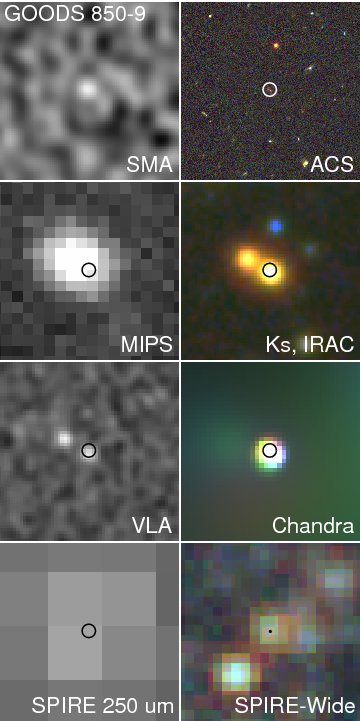}
\includegraphics[width=3.6in]{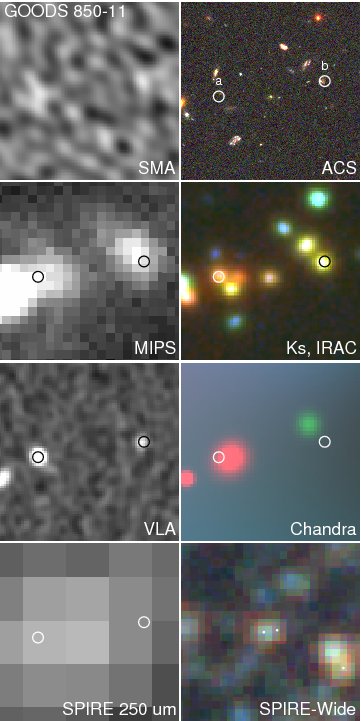}
}
\caption{CONT.}
\end{figure*}

\begin{figure*}[tbh]
\centerline{
\includegraphics[width=3.6in]{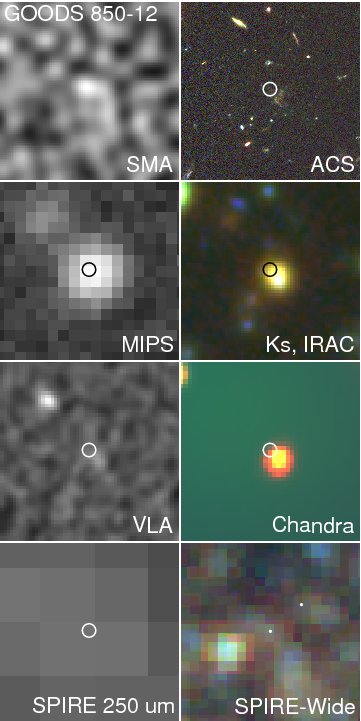}
\includegraphics[width=3.6in]{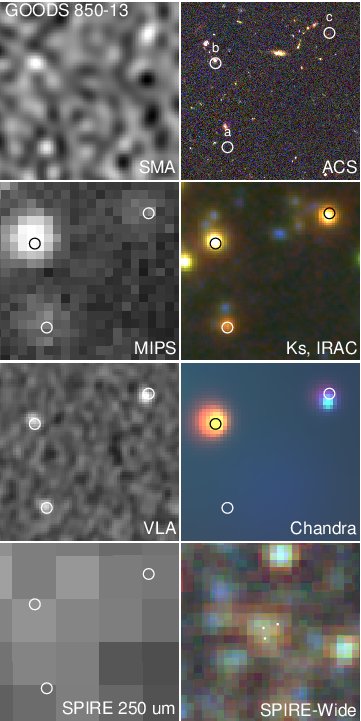}
}
\caption{CONT.}
\end{figure*}

\newpage
\begin{figure*}[tbh]
\centerline{
\includegraphics[width=3.6in]{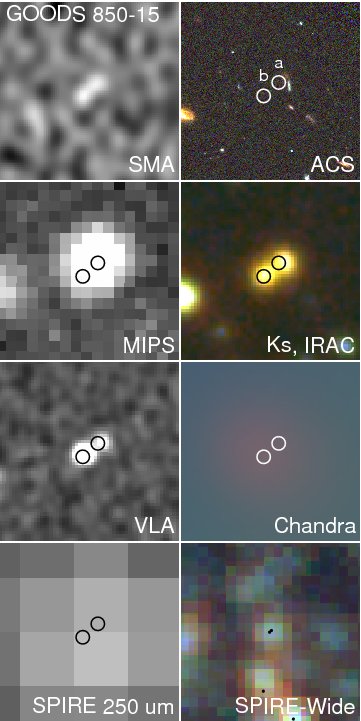}
\includegraphics[width=3.6in]{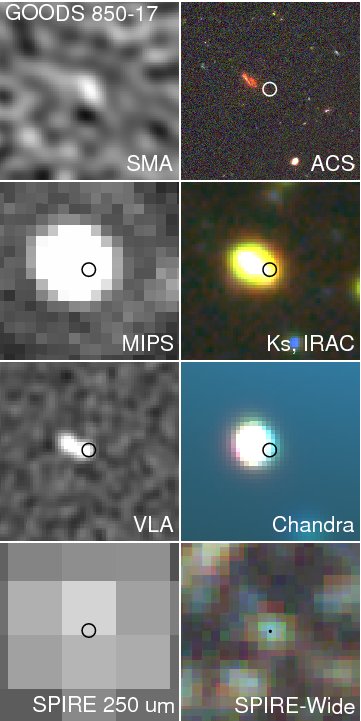}
} 
\caption{CONT.}
\end{figure*}

\end{document}